\documentclass[reprint,aps,pra,eqsecnum, superscriptaddress]{revtex4-2}

\usepackage{amsfonts,amsmath,amssymb}%
\usepackage{bm}
\usepackage{bbm}
\usepackage{booktabs}
\usepackage{comment}
\usepackage{dcolumn}
\usepackage{enumitem}
\usepackage{float}
\usepackage{graphicx}
\graphicspath{{images/}}
\usepackage{epstopdf} 
\usepackage{ifpdf}
\usepackage{lineno,amsfonts}
\usepackage{lipsum}
\usepackage{mathrsfs}
\usepackage{mathtools}
\usepackage{multirow}
\usepackage{natbib}
\usepackage{physics}
\usepackage{qcircuit}
\usepackage{setspace}
\usepackage[table,xcdraw,dvipsnames]{xcolor}
\usepackage{upgreek}

\usepackage{hyperref} 
\hypersetup{colorlinks,
citecolor=blue,
linkcolor=blue,
urlcolor=black,}
\usepackage{cleveref} 

\def\htsg{\hat{\sigma}}
\def\E{{\cal E}}

\def\cti{{\cos\left( \tfrac{ \omega t_i}{2}  \right) }}

\def\sti{{\sin\left( \tfrac{ \omega t_i}{2}  \right) }}

\def \hQ{\hat{Q}}
\def \hR{\hat{R}}
\def \hS{\hat{S}}
\def\one{\hat{\mathbbm{1}}}

\newcommand{\nexpval}[1]{\langle #1 \rangle}

\DeclareMathSymbol{\smin}{\mathbin}{AMSa}{"39}

\allowdisplaybreaks

\newcommand\beq{\begin{equation}}
\newcommand\eeq{\end{equation}}
\newcommand\bea{\begin{eqnarray}}
\newcommand\eea{\end{eqnarray}}

\begin{document}

\title{Detecting violations of macrorealism when \\the original Leggett-Garg inequalities are satisfied}

\author{Shayan Majidy} 
\email{smajidy@uwaterloo.ca}
\affiliation{Institute for Quantum Computing and Department of Physics and Astronomy, University of Waterloo, Waterloo, Ontario N2L 3G1, Canada} 
\author{Jonathan J. Halliwell}
\affiliation{Blackett Laboratory, Imperial College, London SW7 2BZ, United Kingdom}
\author{Raymond Laflamme} 
\affiliation{Institute for Quantum Computing and Department of Physics and Astronomy, University of Waterloo, Waterloo, Ontario N2L 3G1, Canada} 
\affiliation{Perimeter Institute for Theoretical Physics, Waterloo, Ontario N2L 2Y5, Canada} 
\date{\today}

\begin{abstract}
Macroscopic realism (MR) is the notion that a time-evolving system possesses definite properties, irrespective of past or future measurements. Quantum mechanical theories can, however, produce violations
of MR. Most research to date has focused on a single set of conditions for MR, the Leggett-Garg inequalities (LGIs), and on a single data set, the ``standard data set", which consists of single-time averages and second-order correlators of a dichotomic variable Q for three times. However, if such conditions are all satisfied, then where is the quantum behavior? In this paper, we provide an answer to this question by considering expanded data sets obtained from finer-grained measurements and MR conditions on those sets. We consider three different situations in which there are violations of MR that go undetected by the standard LGIs. First, we explore higher-order LGIs on a data set involving third- and fourth-order correlators, using a spin-1/2 and spin-1 system. Second, we explore the pentagon inequalities (PIs) and a data set consisting of all possible averages and second-order correlators for measurements of Q at five times. Third, we explore the LGIs for a trichotomic variable and measurements made with a trichotomic operator to, again, identify violations for a spin-1 system beyond those seen with a single dichotomic variable. We also explore the regimes in which combinations of two and three-time LGIs can be satisfied and violated in a spin-1 system, extending recent work. We discuss the possible experimental implementation of all the above results.
\end{abstract}

\maketitle

\section{Introduction}\label{sec:Intro}

Detecting quantum behavior is necessary for studying the persistence of quantum coherence to the macroscopic domain \citep{friedman2000quantum}, and for identifying quantum effects for the development of new technologies \citep{ladd2010quantum, nielsen2002quantum}. The first tests for detecting such behavior \citep{bell1964einstein} required an entangled bipartite system that is sufficiently separated to rule out communication before measurement \citep{salart2008testing, palacios2010experimental}.
These tests have practical limitations, due to the distance needed, and theoretical limitations, since not all quantum systems are entangled. Leggett and Garg addressed these limitations by proposing a test that does not require an entangled system \citep{leggett1985quantum, leggett2008realism}. In doing so, they defined \textit{macroscopic realism} (MR) - the notion stating that a time-evolving system possesses definite properties irrespective of past or future measurements (see Ref.~\citep{timpson2013quantum} for a discussion on different types of MR). They derived a set of conditions, the original Leggett-Garg inequalities (LGIs) \citep{emary2013leggett}, which if violated would imply a violation of MR, and thus, non-classical behavior.

\begin{figure}
    \centering
    \includegraphics[width=2.25in]{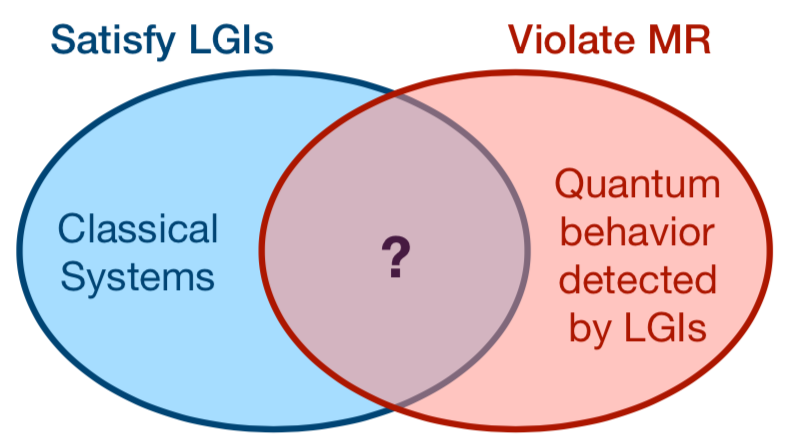}
    \caption{Classical systems that satisfy the LGIs and non-classical systems that violate the LGIs have been well-studied \citep{knee2016strict, goggin2011violation, formaggio2016violation, lambert2016leggett, zhou2015experimental, ku2020experimental, majidy2019exploration}. However, systems that satisfy the LGIs and violate MR have not. In this paper, we identify {experimentally testable cases of such systems. 
    }
    }
    \label{fig:venn_diagram}
\end{figure}

Leggett and Garg's conditions are defined for a particular data set, which we will refer to as the \textit{standard data set}. The standard data set consists of measurements made with a dichotomic observable, {$Q = \pm 1$}, at three times, $t_i$, in which the expectation value at different times, $\nexpval{Q_i} \equiv \nexpval{Q(t_i)}$, and the second-order correlators, $C_{ij} = \nexpval{Q_iQ_j}$, for each pair of times are measured. The three-time LGIs (LG3s) \citep{leggett1985quantum}:
\begin{align}
    1 + s_1s_2C_{12} + s_2s_3C_{23} + s_1s_3C_{13} &\geq 0, \label{eqn:LG3_1}
\end{align}
where $s_i = \{\pm 1\}$ are the measured outcome at $t_i$, provide necessary conditions to satisfy for MR for the standard data set. These conditions are made sufficient for the standard data set if one includes the two-time LGIs (LG2s) \citep{halliwell2019necessary, halliwell2017comparing, halliwell2016leggett, majidy2019exploration}:
\begin{equation}
1 + s_i \nexpval{Q_i} + s_j \nexpval{Q_j} + s_i s_j C_{ij} \geq 0 \label{eq:LG2},
\end{equation}
for each pair of times $ij = 12,23,13$. We will refer to the LG2s and LG3s together as the LGIs.


In this paper we address a limitation in the LGIs by asking the question - Do systems exist which satisfy the LGIs, Eq.~\eqref{eqn:LG3_1} and \eqref{eq:LG2}, but violate MR? If so, what are they? The standard data set provides a very coarse-grained description of a physical system. When the LGIs are satisfied, no quantum behavior is detected. However, if the system does in-fact demonstrate quantum behavior, what else do we need to measure to find it? A natural avenue to explore then are measurements of expanded data sets that provide a finer-grained description of the system. It is both timely and important to address these questions given the LGIs growing use as indicators of quantum behavior in a number of fields \citep{lambert2010distinguishing, li2012witnessing, lambert2013quantum, friedenberger2017quantum}.

We, therefore, investigate {a number of different previously proposed} conditions for MR which go beyond the standard data set in three ways: they include third and fourth-order correlators \citep{halliwell2019leggett}, four or five measurements times \citep{avis2010leggett, halliwell2019fine}, and many-valued variables \citep{halliwell2020conditions}. These conditions will be considered in conjunction with the LG2s and LG3s for the standard data set. Our aim is to explore these sets of conditions for wide parameter ranges for experimentally measurable systems 
that will be relevant to future experimental tests. The regimes that we are interested in finding are those in which the various conditions for the expanded data sets are violated but the LG2s and the LG3s for the standard data set are satisfied. These regimes will demonstrate the aforementioned quantum behavior not detected by the LG conditions involving the standard data set.

From a quantum-mechanical perspective, violations of any MR condition signal the presence of sufficiently large interference terms. This perspective has been analyzed in some detail in Ref.~\citep{halliwell2020conditions} (See also  \citep{kauten2017obtaining, hofer2017quasi, dressel2015weak, pan2020interference}). For the LG2s, the interference terms are proportional to the difference between $ \langle Q_i \rangle$ and $\langle Q_i^M \rangle $, where the latter denotes the average of $Q_i$ in the presence of an earlier measurement that has been summed out. Similarly for the LG3s, the interference terms are proportional to the difference between $C_{ij}$ and $C_{ij}^M$, where $M$ denotes the presence of an earlier or intermediate measurement. The LGIs are the conditions that these interference terms are sufficiently small. {There is also a link here to no-signaling in time (NSIT) conditions -- the requirement that the probabilities for certain sets of sequential measurements are unaffected by the presence of  earlier or intermediate measurements \citep{halliwell2019necessary, halliwell2017comparing, kofler2013condition, clemente2015necessary}. We will discuss this in more detail in what follows.}

The MR conditions for extended data that we explore can be violated in two ways. First, the violations arise from the presence of new types of interference terms that are not seen with dichotomic measurements at two and three times. Second, they arise from the MR condition for the extended data sets depending on the same interferences terms as the LGIs, but in a different manner.

In Sec.~\ref{sec:higher_correlators}, we consider data sets that include the third- and fourth-order correlators, $D_{123} = \nexpval{Q_1Q_2Q_3}$ and $E_{1234} = \nexpval{Q_1Q_2Q_3Q_4}$ \citep{halliwell2019leggett}. We give the MR conditions for the data sets including these higher order correlators and determine the largest possible violation quantum theory can produce. We identify regimes in which the higher order conditions are violated but the LGIs 
are satisfied, for both a spin-$\frac{1}{2}$ and spin-1 system.

In Sec.~\ref{sec:the_PIs}, we consider a data set consisting of the averages and second-order correlators for five times. The usual LGIs will involve different cycles of five correlators. However, there exist conditions, called the pentagon inequalities (PIs) \citep{avis2010leggett, halliwell2019fine},  which contain the full set of ten correlators (Mathematically these conditions are related to polytope constructions, see for example Ref.~\citep{deza2009geometry,avis1978some}). We find that the PIs can be violated when the LG3s are satisfied for a spin-${\frac{3}{2}}$ system and when the LGIs are satisfied for a four- or five-level system.

In Sec.~\ref{sec:many_valued_variables}, we consider data sets involving measurements of a trichotomic operator on a three-level system. The MR conditions explored here are the two and three-time LG conditions for many-valued variables derived in Ref.~\citep{halliwell2020conditions}. Using a spin-1 system, we find that these conditions for a trichotomic variable detect violations of MR that go undetected by the standard LGIs for a single dichotomic variable. We also take advantage of the spin-1 formalism we have set up to explore the relationship between two, three, and four-time LGIs, building on the exploration for the spin-$\frac{1}{2}$ case carried out in Ref.~\citep{majidy2019exploration}. This falls outside the main theme of this paper so this material is described in the supplementary materials section. Finally, we summarize our results in Sec.~\ref{sec:conclusion}. Table \ref{tbl:overview} provides a brief overview of the data sets that are explored. These data sets and MR conditions may be unfamiliar to readers who are familiar with the usual LGIs. So, for perspective, in Table \ref{tbl:conditions}, in Appendix \ref{app:mrzoo}, we give a comprehensive list of a wide range of data sets and their MR conditions. {Furthermore, in Appendix \ref{app:nsit} we sketch the connection with the NSIT conditions aluded to above, which give an alternative way of identifying the interference terms responsible for MR violations.}

\begin{table}
\setlength{\tabcolsep}{5.85pt}
\begin{tabular}{@{}lcccccc@{}}
\toprule
 & Correlator & Times & Lvls & Opr &  Ref.\\ \midrule
Standard & 1st - 2nd & 3 & --  & Di. & \citep{majidy2019exploration, halliwell2019necessary, halliwell2017comparing, halliwell2016leggett}\\
Sec.~\ref{sec:higher_correlators} & 1st - 4th & 3-4 & 2-3 & Di. & \citep{halliwell2019leggett}\\
Sec.~\ref{sec:the_PIs} & 1st - 2nd & 5 & 2-5  & Di. & \citep{halliwell2019fine,avis2010leggett} \\
Sec.~\ref{sec:many_valued_variables} & 1st - 2nd & 3 & 3 & Tri. & \citep{halliwell2020conditions} \\ \bottomrule
\end{tabular}
\caption{The data sets studied in each section, and the standard data set for comparison. The tables list the order of correlators that are measured (Correlators), the number of measurement times (Times), the systems number of levels (Lvls), and whether the type of operator used for the measurements (Opr) is dichotomic (Di.) or trichotomic (Tri.).
}
\label{tbl:overview}
\end{table}

To maintain the focus on the main theme of the paper, many technical details have been relegated to a {further series of appendices (Appendix \ref{app:luders} to \ref{sec:Neq4b})}. Furthermore, the details of the numerical searches used to find the regimes of interest in the paper are provided in the Supplemental Material \cite{supplementary}. Each section instead presents examples found via these searches.

{Although the LG approach \cite{leggett1985quantum, leggett2008realism} was originally motivated by a desire to investigate quantum coherence at the macroscopic scale (and there have been some interesting recent developments in this \cite{knee2016strict}), many theoretical and experimental works on the LG inequalities do not venture very far beyond the microscopic and the present work is no exception in this. However, the types of MR tests discussed here are readily extended to systems which could be potentially macroscopic, e.g. the systems described by continuous variables discussed in two recent papers \cite{bose2018nonclassicality,halliwell2021leggett}, and this is an interesting avenue for future work. Furthermore, since many microscopic systems can have wide parameters ranges in which certain LG inequalities are satisfied,  LG tests have often been usefully employed in such systems to determine where the truly quantum behavior lies (and this is indeed what we do here).}

{A second key aspect of LG tests not emphasized so far is the importance of non-invasive measurements in the measurements of the correlators (at two times or more), in order to rule out the possibility that MR violations can be explained purely using the invasive effects of measurements \cite{montina2012dynamics, yearsley2013leggett}. Non-invasiveness is typically accomplished in LG tests using ideal negative measurements \cite{leggett1985quantum, leggett2008realism}. Refinements of the ideal negative measurement procedure, applicable to more complicated situations like higher order correlators, have been proposed \cite{halliwell2019leggett}. Different types of non-invasive measurement procedures have also been proposed \cite{halliwell2016leggett_ctvm} and experimentally implemented \cite{majidy2019exploration}. In what follows, we are assuming that any experimental tests of the MR conditions discussed here will employ some of these non-invasive measurement techniques.}

\section{Data sets with higher-order correlators} \label{sec:higher_correlators}

\subsection{Conditions for MR - higher-order LGIs}\label{sub2:conditions}

The first expanded data set that we consider includes higher-order correlators. Conditions for MR involving all possible correlators for {$n$-measurement times} were proposed in Ref.~\citep{halliwell2019leggett} (see also Ref.~\citep{pan2018quantum, bechtold2016quantum}). The candidate joint probability for performing a measurement of operators $Q_1, Q_2, \hdots, Q_n$ and returning the outcome $s_1, s_2, \hdots, s_n$ must exist for a macrorealistic theory \citep{halliwell2019leggett} and has the form
\begin{equation}
    p(s_1, s_2, \hdots, s_n) = \frac{1} {2^n} \langle \prod_{i=1}^n  (1 + s_i Q_i) \rangle \ge 0.
    \label{eq:ntimes}
\end{equation}
Eq.~\eqref{eq:ntimes}, for all outcomes, forms a set of necessary and sufficient conditions for MR for the data set in which all possible correlators are fixed. We will refer to these conditions as the \textit{higher order LGIs}.

We consider the $n=3$ and $4$ cases of the higher order LGIs. For $n=3$, measurements are done in seven separate experiments to determine all possible correlators for up to three times. These values are then used to test the 3rd-order LGIs which, from Eq.~\eqref{eq:ntimes}, take the form,
\begin{align}
    p(s_1,s_2,s_3) =& \frac{1}{8} (1 + s_1 \langle Q_1 \rangle + s_2 \langle Q_2 \rangle +s_3 \langle Q_3 \rangle \nonumber \\ &+ s_1 s_2 C_{12}
    + s_2 s_3 C_{23} + s_1 s_3 C_{13}  \nonumber \\ &+ s_1 s_2 s_3 {D_{123}} )\ge 0 \label{3rdOrderLGIs}.
\end{align}
A similar process for the 4th-order LGIs gives,
\begin{align}
     p(s_1,s_2,s_3,s_4) =& \frac{1}{16} (1 + \sum_i^4 s_i \langle Q_i \rangle  + \sum_{i<j}^4 s_i s_j C_{ij}  \nonumber \\
    & + \sum_{i<j<k}^4 s_i s_j s_k {D_{ijk}} \nonumber \\
    & + s_1s_2s_3s_4E_{1234} )\ge 0 \label{4thOrderLGIs}.
\end{align}
Note that the correlators take a different form in a MR model and a quantum mechanical model. The same symbol however, e.g. $C_{12}$, is used for both models. To detect a violation of MR, the calculations must be done using the form according to quantum mechanics. However, what an experimentalist will measure in a lab is independent of this choice.

For measurements of a dichotomic operators, the LGIs, in quantum theory, have a lower bound of $-\frac{1}{2}$, the so-called L\"uders bound \citep{budroni2014temporal, dakic2014density, wang2017enhanced, kumari2018violation, pan2018quantum}. Similarly, the 3rd-order LGIs have lower bound of $-1$ and the 4th-order LGIs have lower bound of $-2$. A proof of the L\"uders bound for the 3rd and 4th-order LGIs is given in Appendix \ref{app:luders}.

When performing projective measurements, the quantum analog of the two-time correlator is \citep{fritz2010quantum}
\begin{align}
    C_{ij} &=  \frac{1}{2}\nexpval{\acomm{\hat{Q}_i}{\hat{Q}_j}}. \label{eq:form_of_Cij}
\end{align}
The same procedure used to determine Eq.\eqref{eq:form_of_Cij} can, as shown in in Appendix \ref{app:corr_forms}, be used to find the forms of the next two higher-order correlators. These are
\begin{align}
    D_{ijk} =& \frac{1}{4}\nexpval{\acomm{\hat{Q}_i}{\acomm{\hat{Q}_j}{\hat{Q}_k}}} \label{eq:form_dijk}\\
    E_{ijkl} =& \frac{1}{8}\nexpval{\acomm{\hat{Q}_i}{\acomm{\hat{Q}_j}{\acomm{\hat{Q}_k}{\hat{Q}_l}}}}, \label{eq:form_eijkl}
\end{align}
and they highlight a pattern in the correlators form.

It is useful to clarify here the difference between, for example, the 3rd-order LGIs and the LGIs for the standard data set. The LGIs are necessary and sufficient conditions that there exist \textit{some} joint probability distribution $p(s_1,s_2,s_3)$ that matches the data set consisting of the $\langle Q_i \rangle$ and $C_{ij}$ \citep{halliwell2019necessary, halliwell2017comparing, halliwell2016leggett}. That is, there exists \textit{some} value for the unmeasured third order correlator for which $p(s_1,s_2,s_3)$ is non-negative. By contrast, the 3rd-order LGIs includes a measurement of the third-order correlators and then tests whether $p(s_1,s_2,s_3)$ is non-negative.

The third order LGIs {detect for both the presence of interference terms already present in the LG2s and LG3s (but here in a new combination), and also for the presence of a new interference term not seen in the lower order tests. This is described in Appendix \ref{app:nsit}.}

Although the LG2s and LG3s may be satisfied or violated independently, their relationship with the 3rd-order LGIs is more restricted. If the 3rd-order LGIs hold then both the LG2s and LG3s must hold since the 3rd-order LGIs explicitly provide an underlying joint three-time probability. However, the LG2s and LG3s may hold even if the 3rd-order LGIs are violated. The same is true for the LG2s and LG3s relationship with the 4th-order LGIs.

In this section, we perform three different searches to find the regimes in which
\begin{enumerate}
    \item The LG2s and LG3s are satisfied but the 3rd-order LGIs are violated.
    \item  The LG2s and LG3s are satisfied but the 4th-order LGIs are violated.
    \item All the 3rd-order conditions are satisfied and the 4th-order LGIs are violated.
\end{enumerate}
In Sec.~\ref{sub2:analytical} we present an analytical solution for finding the first regime for a spin-$\frac{1}{2}$ system. In Sec.~\ref{sub2:numerical} we use a numerical search to find all three regimes for a spin-$\frac{1}{2}$ system. The spin-$\frac{1}{2}$ system is very simple and the form of the correlators is not representative of more general examples. We, therefore, repeat the first two searches for the more complicated case of a spin-$1$ system.

\subsection{Analytic examples}\label{sub2:analytical}

Consider a spin-$\frac{1}{2}$ system that is measured with a dichotomic operator $\hQ = \vec{c}\cdot \vec{\sigma}$, where $\vec{c}$ is a unit vector and $\vec{\sigma}$ is the Pauli vector, that is in the initial state
\begin{equation}
    \rho = \frac{1}{2} (I + \Vec{v}\cdot \Vec{\sigma}), \label{eq:sec2_initial}
\end{equation}
where $\vec{v} \cdot \vec{v} \leq 1$ and the equality holds for pure states. We define $\vec{c}_i$ such that $\hQ_i = e^{i\hat Ht}\hQ e^{-i\hat Ht} \equiv \vec{c}_i \cdot \vec{\sigma}$, so $\vec{c}_i$ is a unit vector and a function of $\vec{c}$, $\hat{H}$ and $t$. We find then, 
\begin{align}
\nexpval{\hat{Q}_i} &= \vec{c}_i\cdot \vec{v}\\
C_{ij} &= \vec{c}_i \cdot \vec{c}_j\\
D_{ijk} &= (\vec{c}_j \cdot \vec{c}_k) (\vec{c}_i \cdot \vec{v})\\
E_{ijkl} &= (\vec{c}_i \cdot \vec{c}_j)(\vec{c}_k \cdot \vec{c}_l)
\end{align}
We find then that, in the spin-$\frac{1}{2}$ model, the higher-order correlators are products of the lower ones, $D_{ijk} = \nexpval{Q_i}C_{jk}$ and $E_{ijkl} = C_{ij}C_{kl}$ (This simple form does not persist to the spin-1 case, as we will show).

The MR conditions for three times can thus be written in simpler forms. The LG2s can then be written as
\begin{align}
1  + s_i \vec{c}_i\cdot\vec{v} + s_j \vec{c}_j\cdot\vec{v} + s_is_j \vec{c}_i\cdot\vec{c}_j \geq 0 \label{eq:vecproof1}\\
 (\vec{v} + s_i \vec{c}_i + s_j \vec{c}_j)^2 \geq 1, \label{eq:vecproof2}
\end{align}
for each pair of times and the LG3s can be written as
\begin{align}
   1 + s_1 s_2 \vec{c}_1 \cdot \vec{c}_2 + s_2 s_3 \vec{c}_2 \cdot \vec{c}_3 + s_1 s_3 \vec{c}_1\cdot \vec{c}_3 \geq 0 \label{eq:vecproof3}\\
    (s_1 \vec{c}_1 + s_2 \vec{c}_2 + s_3 \vec{c}_3)^2 \geq 1. \label{eq:vecproof4}
\end{align}
Note that Eqs.~\eqref{eq:vecproof1} and \eqref{eq:vecproof1} are the same equation written in two different ways and likewise for Eqs.~\eqref{eq:vecproof3} and \eqref{eq:vecproof4}. The second form in both cases being a more convenient way for spotting regimes in which the inequalities are satisfied. Furthermore, the 3rd-order LGIs can be written as,
\begin{align}
&1  + s_1 \vec{c}_1\cdot\vec{v} + s_2 \vec{c}_2\cdot\vec{v} + s_3 \vec{c}_3\cdot\vec{v} 
\nonumber \\ &
+ s_1s_2 \vec{c}_1\cdot\vec{c}_2 + s_1s_3 \vec{c}_1\cdot\vec{c}_3  +  s_2s_3 \vec{c}_2\cdot\vec{c}_3 \nonumber\\ &
+ s_1s_2s_3(\vec{c}_1\cdot\vec{v})(\vec{c}_2\cdot\vec{c}_3)  \geq 0 .
\end{align}
Note that this equation follows from Eq.~\ref{eq:ntimes} and thus represent the three time probability distribution up to an overall factor of $\frac{1}{8}$.

We now present a parameter choices that satisfies the LG2s and LG3s but violates the 3rd-order LGIs. First, we choose $ \vec{c}_2 = - \vec{c}_1$, so that two of the LG3s reduce to $ \vec{c}_3 \cdot \vec{c}_3 \ge 1$ and the other two reduce to $ 1 \pm \vec{c}_1 \cdot \vec{c}_3 \ge 0 $. These conditions always hold for unit vectors. We find that the LG2s also hold for  $t_1,t_2$. Second, we choose $\vec{c}_3 = - \vec{v}$, so that the LG2s for $t_2,t_3$ and $t_1,t_3$ are all satisfied and the third order condition for the $s_1 = s_2 = s_3 = 1 $ case is
\begin{equation}
    -1 + ( \vec{c}_1 \cdot \vec{v} )^2 \ge 0, \end{equation}
which is generally violated. Hence the desired regime is readily found.

We now use this example to illustrate the discussion above concerning the relationship between the third order conditions and conditions involving only the LG2s and LG3s. With the above choice of $\vec{c}_2$ and $\vec{c}_3$, the three-time candidate probability is
\begin{align}
    p(s_1,s_2,s_3) &= \frac{1}{8} \Big( 1 - s_3 - s_1 s_2    + \big[ s_1 (1-s_3) \nonumber\\
    & - s_2 (1-s_3) \big] (c_1.v) + s_1 s_2 s_3 D_{123} \Big)
\end{align}
where we keep the third order correlator $D_{123}$ general for the moment. As discussed above, if both the LG2s and LG3s are satisfied, as they are in this case, then the fact that these are sufficient conditions for MR means that there must exist some choice of $D_{123}$ for which $p(s_1,s_2,s_3)$ is non-negative. It is not hard to show that the only value doing the job is $D_{123}=1$. For this value, $p(s_1,s_2,s_3)$ is identically zero for $s_3=+1$, and also for $s_3=-1$ with $s_1=s_2$. For the remaining case $s_3=-1$ and $s_1=-s_2$, non-negativity is equivalent to requiring that
\begin{equation}
1 + s_1(c_1.v)  \ge 0,    
\end{equation}
which is clearly true. On the other hand, if we take the larger data set in which $D_{123}$ is determined by  measurement, not freely chosen, we find the value $D_{123} = (c_1.v)^2$ in this case, which can take values other than $1$, and the third order condition is then violated even though the LG2s and LG3s are satisfied.

\subsection{Higher-order LGI violations in spin-\texorpdfstring{$\frac{1}{2}$}{1/2} and spin-1 models}\label{sub2:numerical}

We now find all three regimes of interest outlined in Sec.~\ref{sub2:conditions} via a numerical search. We consider here a more specific spin-$\frac{1}{2}$ model, one in which the Hamiltonian is $\omega \htsg_x/2$ and the dichotomic operator is $\htsg_z$. 
We thus find,
\begin{align}
    \nexpval{\hQ_i} &= v_z \cos(\omega t_i) + v_y \sin(\omega t_i)\\
    C_{ij} &= \cos(\omega(t_j - t_i)) \label{eq:Cij_simp_form}.\\
    D_{ijk} &= \cos(\omega(t_k-t_j))(v_z \cos(\omega t_i) + v_y \sin(\omega t_i)) \\
    E_{ijkl} &= \cos(\omega(t_j - t_i)) \cos(\omega(t_l - t_k)).
\end{align}

\begin{figure}
    \centering
    \includegraphics[width=\linewidth]{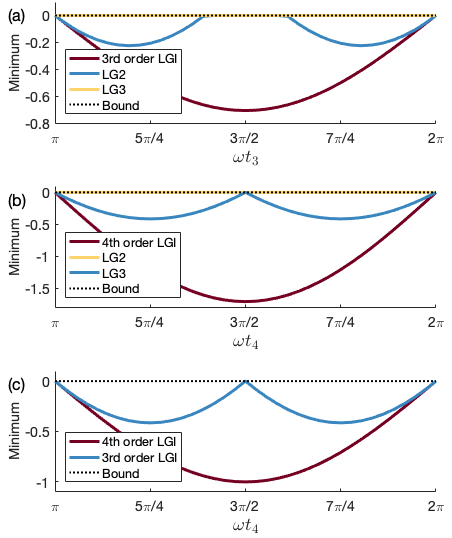}
    \caption{Spin-$\frac{1}{2}$ systems that violate higher-order LGIs but not the LG2s and LG3s. (a) The minimum values of the LG2s, LG3s and 3rd-order LGIs as a function of $\omega t_3$ for the initial state defined by $\vec{v} = (1,1,0)/\sqrt{2}$ (see Eq.~\eqref{eq:sec2_initial}) for $\omega t_1 = 0,$ and $\omega t_2 = \pi$. The plot for the LG3s and the bound are overlapping. (b) The minimum values of the LG2s, LG3s and 4th-order LGIs as a function of $\omega t_4$ for the initial pure state defined by $\vec{v} = (1,0,1)/\sqrt{2}$ for $\omega t_1 = 0, \omega t_2 = \tfrac{\pi}{2}$, and $ \omega t_3 =\pi$. The plot for the LG2s and the bound are overlapping. (c) The minimum values of the 3rd-order LGIs and 4th-order LGIs as a function of $\omega t_4$ for the initial pure state defined by $\vec{v} = (1,0,0)$ for $\omega t_1 = 0, \omega t_2 = \tfrac{\pi}{2},$ and $\omega t_3 = \pi$.}
    \label{fig:spin_half_higher_correlators}
\end{figure}

Fig.~\ref{fig:spin_half_higher_correlators} presents examples of the three regimes explored in this section. The examples chosen highlight how easily and trivially the LG2s and LG3s are satisfied but the higher-order conditions are violated. For example, for the parameters in Fig.~\ref{fig:spin_half_higher_correlators}a we have that $\langle \hQ_1 \rangle = \langle \hQ_2 \rangle = 0, \langle \hQ_3 \rangle = \tfrac{1}{\sqrt{2}}\sin(\omega t_3), C_{12} = -1, C_{13} = - C_{23} = \cos(\omega t_3),$ and $D_{123} = 0$. The LG2 for $t_1,t_2$ reduces to $1 - s_1s_2 \geq 0$ and is always satisfied. The remaining two LG2s become
\begin{align}
    1 &+ \frac{s_3}{\sqrt{2}}\sin(\omega t_3) + s_1s_3 \cos(\omega t_3) \geq 0,\\
    1 &+ \frac{s_3}{\sqrt{2}}\sin(\omega t_3) - s_2s_3 \cos(\omega t_3) \geq 0,
\end{align}
which are always satisfied for $2\pi - 2\arctan(\sqrt{2}) \leq \omega t_3 \leq  2\pi - 2\arctan(\frac{1}{\sqrt{2}})$, which is a region centred at $\frac{3\pi}{2}$. The LG3s for these correlators become
\begin{equation}
    1 - s_1s_2 + s_3\cos(\omega t_3)(s_1 - s_2) \geq 0,
\end{equation}
which equals either $0$, $1+\cos(\omega t_3)$, or $1 - \cos(\omega t_3)$, all of which are always satisfied. The 3rd-order LGIs become
\begin{align}
1 - s_1s_2 + \frac{s_3}{\sqrt{2}}\sin(\omega t_3) + s_3\cos(\omega t_3)(s_1 - s_2) \geq 0.  
\end{align}
Which for all $s_i = +1$ has a local minimum which is negative at $\omega t_3 = 3\pi/2$. We thus find a regime in which the 3rd-order LGIs are violated and the LG2s and LG3s are satisfied. This regime was found for a system which can easily detected with a nuclear magnetic resonance (NMR) spectrometer. LGIs tests have been done with NMR for different two-level \citep{athalye2011investigation, majidy2019violation} and three-level systems \citep{katiyar2017experimental}.

\section{Data sets with additional measurement times} \label{sec:the_PIs}

\subsection{Conditions for MR - the pentagon inequalities}

The second expanded data set that we consider involves measurements of the averages and second-order correlators at more than three times. These conditions were first proposed in Ref.~\citep{avis2010leggett} for the $n=5$ case and then generalized for $n$ measurements in Ref.~\citep{halliwell2019fine} to the following conditions,
\begin{equation}
    n + 2 \sum_{i>j}^n s_i s_j C_{ij} \ge 
    \begin{cases}
    1   &\mbox{if} {\ n \  {\rm odd} },  \\ 
    0  & \mbox{if} {\ n \ { \rm  even} }.
    \end{cases}
    \label{neweq}
\end{equation}
For $n=3$, Eq.~\eqref{neweq} takes the form of the LG3s. For $n=4$, Eq.~\eqref{neweq} takes the form of
\begin{equation}
    2 + \sum_{i>j}^n s_i s_j C_{ij} \ge 0.\label{Av2}
\end{equation}
However, Eq.~\eqref{Av2} can be written as an average of four sets of LG3s in this case. (Note that these conditions for the $n=4$ case involve all six of the possible correlators, so are different to the usual four-time LG inequalities, which involve a cycle of four correlators chosen from the six). For the $n=5$ case, Eq.~\eqref{neweq} again takes the form of Eq.~\eqref{Av2}. However, as shown in Ref.\citep{avis2010leggett}, these conditions cannot be written as averages of LG3s in this case. Thus, it is only for $n \ge 5$ that conditions of this form could be stronger than the LG3s. 

For the $n=5$ case, Eq.~\eqref{Av2} is known as the pentagon inequalities (PIs) \citep{avis2010leggett}. For the data set that consists of the five averages and a cycle of five $C_{ij}$, the necessary and sufficient conditions for MR are the LG2s and five-time LGIs (LG5s) for the given cycle. If the data set expands to all ten $C_{ij}$, the necessary and sufficient conditions are the set of all possible LG2s, LG3s, and PIs \citep{halliwell2019fine}.

Regimes in which the LG3s and PIs were violated/satisfied was first explored in Ref.~\citep{avis2010leggett}. It was found to be impossible to violate the PIs and satisfy the LG3s for a spin-$\frac{1}{2}$ system with equal time spacing between measurements. A modified situation was then investigated in which a dephasing gate was applied as an intermediate measurement between certain pairs of times in some, but not all, experiments. This modifies the correlators in some experiments and yields a PI violation with the LG3s satisfied. This is a natural avenue to investigate on the grounds that such a procedure makes no difference from the macrorealistic point of view. However, the decoherence procedure was applied differently across experiments - for example, it was applied in the measurement of $C_{13}$ but not $C_{14}$. It would clearly be more desirable to avoid this feature.
 
A natural question then is - what is it that PI violations are detecting from a quantum-mechanical perspective? 
The PI can be violated when the LGIs are satisfied either from the presence of five-time interference terms, not detected by the LGIs, or by the combination of interferences at two and three times that are not large enough to produce LG2 and LG3 violations {(This is discussed in Appendix \ref{app:nsit})}.

In the remainder of this section, we give two examples of PI violations. The first is found via a numerical search for spin models of spin-$\frac{1}{2}$, $1$ and $\frac{3}{2}$. For the spin-$\frac{3}{2}$ model, we find a significant regime in which the PIs are violated and all LG3s are satisfied. 
However, the LG2s were not satisfied in this case. This is not a significant shortcoming since the LG2s detect interferences which are, in general, independent of the interferences causing PI violations. None-the-less, we then give a second example, using an analytic model for a {five-level system} with a specially constructed Hamiltonian, where the PIs are maximally violated and all possible LG2s and LG3s are satisfied. We then find that a similar construction can be done for a {four-level system}.

\subsection{Form of the Correlators and PI violations in a spin-\texorpdfstring{$\frac{3}{2}$}{3/2} models}\label{sub:PI1}

{Consider a dichotomic variable defined for a system with $N$-levels (note that $N$ should not be confused with $n$, which denotes the number of measurements). A convenient set of such} dichotomic variables to work with are those of the form,
\begin{align}
    \hat Q &=  1 - 2\dyad{A} \
\end{align}
for some state $|A \rangle$, for which there are $N$ choices. (For $N \ge 4$, this is not the only possible form). The time evolution of $\hat{Q}$ then boils down to the time-evolution of $\ket{A}$. Thus, for $\ket{v_i} \equiv e^{ - i \hat{H} t_i} \ket{A}$, we have
\begin{equation}
     \hat Q_i \equiv 1 - 2 \dyad{v_i} 
\end{equation}
The averages and second order correlators are then,
\begin{align}
    \langle \hat Q_i \rangle &=  1 - 2 \abs{\ip{\psi}{v_i}}^2, \label{eq:general_Qi}\\
    C_{ij} &= 1 - 2 \abs{\ip{\psi}{v_i}}^2  - 2 \abs{\ip{\psi}{v_j}}^2 \nonumber \\
    & + 4\Re(\ip{\psi}{v_i}\ip{v_i}{v_j}\ip{v_j}{\psi}),\label{eq:general_Cij}
\end{align}
where $\ket{\psi }$ is the initial state. Eq.~\eqref{eq:general_Cij} is found using $C_{ij} = {\rm Re} \mel{\psi}{\hat Q_i \hat Q_j}{\psi}$ (which follows from Eq.~\eqref{eq:form_of_Cij}).

For experimental feasibility, an analysis with a fixed Hamiltonian is useful. 
We write our Hamiltonian generally as $ \hat{H} = \sum_n  \E_n  \dyad{\E_n }{\E_n }$, where $\E_n$ and $ \ket{\E_n}$ are the eigenvalues and orthonormal eigenvectors. Thus the unit vectors $\ket{v_i}$ which can arise in a physical model are
\begin{equation}
    \ket{v_i} =  \sum_n e^{ - i \E_n t_i}  \ket{\E_n} \ip{\E_n}{A}. \label{eq:visum}
\end{equation}

\begin{figure}
    \centering
    \includegraphics[width=\linewidth]{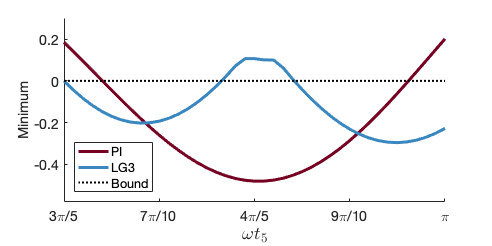}
    \caption{An example of a spin-$\frac{3}{2}$ system which violates the PIs but not the LG3s. The minimum values of the LG3s and PIs are plotted as a function of $\omega t_5$ for the initial pure state $\rho$ defined by $\theta = 9\pi/5$, $\alpha = 9\pi/5$, $\beta = 0$, $\phi_1 = 3\pi/5$, $\phi_2 = 6\pi/5$ and $\phi_3 = 9\pi/5$ (for the parameterization given in Appendix \ref{sec:Neq4b}) with $\omega t_1 = 0$, $\omega t_2 = \pi/5$, $\omega t_3 = 2\pi/5$ and $\omega t_4 = 3\pi/5$. {Note that the LG3 plot seems to change unnaturally around $4\pi/5$. This oddity is a result of the plot being the minimum of many different functions and not from a sparsity of plotted points.}}
    \label{fig:example_LG3_vs_PI}
\end{figure}

The Hamiltonian that we will initially be considering is that of a $N$-level spin system, $\hat{H} =  \omega\sigma^{(N)}_x/2$ \citep{levitt2013spin}, where $\sigma^{(N)}_x$ is the $N$-level Pauli-$x$ matrix. We find that a regime in which the LG3s are satisfied and the PIs are violated does exist for a spin-$\frac{3}{2}$ system, as shown in Fig.~\ref{fig:example_LG3_vs_PI}.

\subsection{Analytic examples in four and five-level systems} \label{sub:PI2}

Our second set of examples consists of analytically tractable models in four and five Hilbert space dimensions. In these models, the state and Hamiltonian are chosen such that all the correlators $C_{ij}$ are equal and all the averages $\langle Q_i \rangle$ are equal. In this case, if the correlators all lie in the range $ -\frac{1}{3} \le C_{ij} < - \frac{1}{5} $, then the PIs are violated and all LG3s are satisfied. The LG2s are then also satisfied if $ \langle Q_i \rangle \ge - \frac{1}{3}$. 

Achieving such a construction is straightforward if we can arrange the values of $\ip{v_i}{v_j}$ to be equal for all $i \ne j$. Geometrically, it is reasonably clear that we can always do this for $N=5$. Consider the five orthogonal basis vectors $\ket{e_i}$, for $i = 1,\hdots 5$. The normalized vector which is at the centre of these basis vectors can be denoted as,
\begin{equation}
\ket{u} = \frac{1}{\sqrt{5}} \sum_{i=1}^5 \ket{e_i}
\end{equation}
and it has an equal overlap with each $\ket{e_i}$. We then define the non-orthogonal vectors $\ket{v_i}$ by rotating each $\ket{e_i}$ towards $\ket{u}$. This may conveniently be written as,
\begin{align}
\ket{v_i} = e^{ i \phi} \cos \theta \ket{e_i} + \frac{1}{2} e^{- i \phi} \sin \theta \sum_{j \neq i} \ket{e_j},
\end{align}
where $\ket{v_i}$ is normalized. We find that for all $i \ne j$,
\begin{equation}
\langle v_i | v_j \rangle = \frac{3}{4} \sin^2 \theta  + \frac{1}{2} \sin (2 \theta) \cos (2 \phi). \label{eq:theoretical_vivj}
\end{equation}
By defining $\alpha \equiv \langle v_i | v_j \rangle$ and choosing our normalized initial state to be $\ket{\psi} = \ket{u}$, we have that,
\begin{equation}
\langle v_i | \psi \rangle = \sqrt{  \frac{ 1 + 4 \alpha}{5} }. \label{eq:theoretical_vipsi}
\end{equation}
By substituting Eq.~\eqref{eq:theoretical_vivj} and  \eqref{eq:theoretical_vipsi} into Eq.~\eqref{eq:general_Qi} and \eqref{eq:general_Cij} we find that for all $i<j$,
\begin{align}
    \nexpval{Q_i} &= \frac{ 3 - 8 \alpha}{5},\\
    C_{ij} &= 1 - \frac{4}{5} ( 1 - \alpha)  (1 + 4 \alpha). \label{Cij5}
\end{align}
We achieve the desired ranges of
 $ -1/3 \leq C_{ij} <  -1/5 $ and $  \langle \hat Q_i \rangle \ge -\frac{1}{3}$ by choosing alpha to be in the range  $[\frac{1}{4}, \frac{1}{2}]$. Here we choose $\alpha = 3/8$, which is achievable according to Eq.~\eqref{eq:theoretical_vivj}.
 For this choice, the PI with all plus signs is violated and reaches the maximum violation of $-\frac{1}{2}$. 
 
This regime is also physically realizable. Consider the unitary evolution,
\begin{equation}
    \hat{U} = \dyad{e_1}{e_2} + \dyad{e_2}{e_3} + \dyad{e_3}{e_4} + \dyad{e_4}{e_5} + \dyad{e_5}{e_1},
\end{equation}
which cycles between $\ket{v_i}$. We can then write $\hat{U} = e^{-i\hat{H}t}$ and solve for $\hat{H}$. Since the eigenstates of $\hat{U}$ are,
\begin{equation}
    \ket{\E_k} = \frac{1}{\sqrt{5}}\sum_i e^{i2\pi k n/5} \ket{e_i},
\end{equation}
we can construct the following time-independent Hamiltonian which implements the desired transformation,
\begin{equation}
    \hat{H} = \sum_k \frac{2\pi k}{5} \dyad{\E_k}.
\end{equation} 
A system with such a Hamiltonian can be experimentally implemented and measured using a programmable quantum device \citep{nielsen2002quantum}. Trapped-ion \citep{bruzewicz2019trapped} or superconducting qubit systems \citep{blais2020quantum} could be used for such a test.

A similar construction also works for $N=4$. Consider the four vectors
\begin{align}
\ket{v_i} &= e^{ i \phi} \cos \theta \ket{e_i}  + \frac{1}{\sqrt{3}} e^{- i \phi} \sin \theta
 \sum_{j \neq i }^5 \ket{e_j} 
\end{align}
for $i = 2,3,4,5$. We find then that the overlap between each of these four vectors is
\begin{equation}
    \alpha = \langle v_i | v_j \rangle = \frac{2}{3} \sin^2 \theta  + \frac{1}{\sqrt{3}} \sin(2 \theta) \cos (2 \phi).
\end{equation}
By choosing our initial state and $\ket{v_1}$ to be,
\begin{equation}
| \psi \rangle =  |v_1 \rangle = \frac{1}{ 2 \sqrt{ 1 + 3 \alpha} } \sum_{i=2}^5 |v_i \rangle,
\end{equation}
we find that
\begin{equation}
\langle v_1 | v_j \rangle = \frac{1}{2} \sqrt{ 1 + 3 \alpha}.
\end{equation}
We find, from Eq.~\eqref{eq:general_Cij}, that the correlators for $i,j = 2,3,4,5$ are
\begin{equation}
C_{ij} = - 2 \alpha + 3 \alpha^2,
\end{equation}
and for $j=2,3,4,5$ are
\begin{equation}
C_{1j} = -\frac{1}{2} + \frac{3}{2} \alpha.
\end{equation}
Both types of correlator take the value $-1/4$ at $\alpha = 1/6$, hence all the LG3s are robustly satisfied and the PIs achieve their maximum violation. From Eq.~\eqref{eq:general_Qi}, we find the single time averages are 
\begin{equation}
\langle Q_i \rangle =  \frac{1}{2} - \frac{3}{2} \alpha = - C_{1i}
\end{equation}
for $i=2,3,4,5$ and $ \langle Q_1 \rangle = -1$. This means that all the LG2s are also satisfied for $\alpha = 1/6$. However, in the case of the LG2s involving $\langle Q_1 \rangle$, they are right on the boundary (i.e. some of them are zero). This means the LG2s are not satisfied as robustly as in the $N=5$ model (although this can often be fixed by including a small amount of decoherence). Nevertheless we have again found the desired regime for a model with $N=4$. A similar approach as in the $N=5$ case can be used to experimentally test this regime with a programmable quantum device.

\section{Data sets with many-valued variables}\label{sec:many_valued_variables}


The final extended data set we consider is the extension from measurements of a single dichotomic variable at two and three times to that of a trichotomic variable in a three-level system.  We therefore consider a system with orthogonal states $\ket{A}, \ket{B}, \ket{C}$. In the standard data set, we would consider measurements, at each time, of just one of the three possible dichotomic variables, $ \hat Q = 1 - 2 \dyad{A}$, $\hat R = 1 - 2 \dyad{B}$, $S = 1 - 2 \dyad{C}$. Each of these variables alone determines whether the system is, or is not, in a given state, but since there are three states, it gives an incomplete account of the full set of possibilities. 

A more complete account is obtained using a trichotomic variable which gives full information as to which state the system is in. An example of such a trichotomic variable is $ |A \rangle \langle A| - | C \rangle \langle C|$, which has eigenvalues $+1, 0 , -1$. However, it turns out to be more convenient to describe the system using the set of three dichotomic variables given above. They clearly satisfy $\hQ + \hR + \hS = -1$, so are not independent which means that any two suffice, but the use of all three permits the relevant LG inequalities to be written down in a form similar to the dichotomic case. 

The extended data set for this system consists of the nine averages $\langle X_i \rangle$ at three times, $i=1,2,3$, and the twenty-seven correlators of the form $ \langle X_i X_j \rangle$, where each $X_i$ may be taken to be $Q_i, R_i$ or $S_i$.
The number of averages and correlators that need to be measured is of course reduced, as a result of the relation above between $Q, R, S$.
The necessary and sufficient conditions for this data set for measurements at two and three times were given in Ref.~\citep{halliwell2020conditions}. They consist of 
the nine LG2s,
\begin{align}
    1 + \nexpval{X_1} +\nexpval{X_2} +\nexpval{X_1 X_2}  & \ge 0, \label{eq:3lvlLG2}
\end{align}
plus two more sets of nine each at times $t_2,t_3$ and $t_1, t_3$,
and the twenty-seven LG3s:
\begin{align}
    1 +  \langle X_1 X_2 \rangle + \langle X_2 X_3  \rangle  + \langle X_1 X_3\rangle \geq 0 \label{eq:3lvlLG3}.
\end{align}

\begin{figure}
    \centering
    \includegraphics[width=\linewidth]{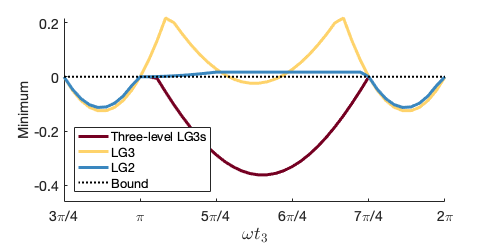}
    \caption{An example of a spin-$1$ system which violate the LG3s for a trichotomic variable but not the dichotomic variables LG2s or LG3s. The minimum values of the LG2s, LG3s, and LG3s for a trichotomic variable are plotted as a function of $\omega t_3$ for the initial pure state $\rho$ defined by $\vec{a} = (0,0,1,2,1,0,0,-1)/3\sqrt{3}$ (see Eq.~\eqref{eq:general_3_level_rho}), with $\omega t_1 = 0$ and $\omega t_2 = 3\pi/4$. {Similar to Fig.~\ref{fig:example_LG3_vs_PI}, the unnatural looking changes in the LG3 and LG2 function are the result of them being the minimum of many different functions.}}
    \label{fig:PIs_LG3s_example}
\end{figure}

By searching over the states of a simple spin-1 model (described in Appendix \ref{app:gen_op_form}) we find a regime in which the LG2s and LG3s for the standard data set with a single dichotomic variable in a three-level system are satisfied, but the above LG2s and LG3s for the extended data set of the trichotomic case can be violated. The dichotomic variable chosen here is $\hQ$. An example of such a regime is presented in Fig.~\ref{fig:PIs_LG3s_example}. Similar to the discussion in Sec.~\ref{sec:higher_correlators}, such a regime can be tested using an NMR system.

We also take advantage of the spin-1 formalism we have set up to carry out an exploration of the relationship between two, three and four time LGIs, building on the exploration for the spin-$\frac{1}{2}$ case carried out in Ref.~\citep{majidy2019exploration}. This however is not in-line with the main theme of the paper and is instead included in the Supplemental Material \citep{supplementary}.

{Like the MR conditions discussed in the previous two sections, there is a story for MR conditions on many-valued variables about interferences and NSIT conditions similar to that discussed in Appendix \ref{app:nsit}. However, this is largely covered in an earlier work \cite{halliwell2020conditions} so will not be repeated here.}

\section{Conclusion} \label{sec:conclusion}

In this paper, we investigated a limitation in the LGIs by asking the question of whether systems exist that satisfy the LGIs but violate MR. We answered this question affirmatively and provided physical examples. 

First, we considered data sets that included the third- and fourth-order correlators, $D_{123} = \nexpval{Q_1Q_2Q_3}$ and $E_{1234} = \nexpval{Q_1Q_2Q_3Q_4}$. We derived the largest possible violations of the MR conditions for these data sets, i.e. the L\"uders bounds. We then identified regimes in which the higher-order conditions were violated but the LGIs were satisfied, for both a spin-$\frac{1}{2}$ and spin-1 system.

Second, we considered data sets consisting of the five averages, $\expval{Q_i}$, and ten second-order correlators, $C_{ij}$, for five times. In this case, the usual LGIs involve different cycles of five correlators. However, the pentagon inequalities (PIs) \citep{avis2010leggett, halliwell2019fine} contain the full set of ten correlators and can detect interference terms that the LGIs cannot. We found that the PIs can be violated when the LG3s are satisfied for a three-level spin system and when the LGIs are satisfied for a five-level system.

Third, we considered data sets involving measurements of a trichotomic operator on a three-level system. The MR conditions explored here were the LG2s and LG3s for a trichotomic variable and LG3s \citep{halliwell2020conditions}. Using a spin-1 system, we found that the LG2s and LG3s for a trichotomic variable detect violations of MR that the dichotomic LGIs cannot.

There exist a few natural extensions of this work. An immediate one, is the experimental tests of the systems identified that violate MR but satisfy the LGIs. A broader extension, is the refinement of the different protocols for detecting quantum behavior, such as protocols for certifying quantum components uniquely and device-independently \citep{lunghi2015self,li2011semi, maity2020self}, to be robust against case where the LGIs are satisfied but MR is still violate. A third extension of this work is identifying the relationship between the different MR conditions which have been discovered over the past decade (those explored in this paper \citep{halliwell2019necessary, halliwell2017comparing, halliwell2016leggett, halliwell2019leggett, halliwell2019fine, avis2010leggett, halliwell2020conditions} as well as others \citep{kofler2013condition,clemente2015necessary}). Such an investigation entails answering the questions of -- Which conditions can be violated independently of others, which conditions are are strictly stronger than other, and what types of interference terms is each measuring?

Aside from these extension, this work also introduces broader questions that are of interest. For one, in this work we demonstrated that with more measurements times and with higher-level systems, it is possible to tease out finer interference terms and detect violations of MR. However, experimentally, both of these extensions will introduce more noise and thus hinder the detection of violations of MR. This then raises the question of whether there exist a noise threshold for MR, in which regardless of the number of measurement times or the levels of the system, a violation of MR is no longer possible. Conversely, we can also ask -- With infinite measurement precision and zero noise, does there always exist a set of measurements which can detect a violation of MR? Both these questions are of clear theoretical interest.

Given the LGIs growing use as indicators of quantum behavior in different fields \citep{lambert2010distinguishing,  li2012witnessing, lambert2013quantum, friedenberger2017quantum}, it has become more important than ever to understand their limitations. Our results highlight where these limitations occur and provide practical solutions for when such limitations are met.

\begin{table*}
\setlength{\tabcolsep}{8pt}
\renewcommand{\arraystretch}{1.2}
\scalebox{0.98}{
\begin{tabular}{@{}c|cccc|cccc|c@{}}
\toprule
\multirow{3}{*}{\textbf{n}} & \multicolumn{4}{c|}{\textbf{Data set}} & \multicolumn{4}{c|}{\textbf{Necessary and sufficient conditions}} & \multirow{2}{*}{\begin{tabular}{@{}c@{}}
    \textbf{Sec.}/ \\ \textbf{Ref.}\end{tabular}}
\\
 & $\nexpval{Q_i}$ & $C_{ij}$ & $D_{ijk}$ & $E_{ijkl}$ & LG2s & LGIs & PIs & Higher-order LGIs
  &   \\ \midrule 
\textbf{2} & All (2) & All (1) & &  & All &  &  & & \citep{halliwell2019necessary, halliwell2017comparing, halliwell2016leggett, majidy2019exploration} \\ \hline
\multirow{2}{*}{\textbf{3}} & All (3) & All (3) & &  & All & All LG3s &  &  & Sec.~\ref{sec:higher_correlators} \\
 & All (3) & All (3) & All (1) &  &  & &  & All 3rd-order & \citep{halliwell2019necessary, halliwell2017comparing, halliwell2016leggett, majidy2019exploration} \\ \hline 
\multirow{3}{*}{\textbf{4}} & All (4) & Cycle of 4 & &  & {LG2s FTC} & LG4s {FTC} &  & &  \\
 & All (4) & All (6) & &  & All & All LG3s &  & &  Sec.~\ref{sec:higher_correlators} \\
  & All (4) & All (6) & All (4) & All (1) &  &  &  & All 4th-order &  \\
 \hline
\multirow{2}{*}{\textbf{5}} & All (5) & Cycle of 5 & &  & {LG2s FTC} & LG5s {FTC} &  & &  \multirow{2}{*}{Sec.~\ref{sec:the_PIs}}\\
 & All (5) & All (10) & &  & All & All LG3s & All & &  \\ \bottomrule
\end{tabular}
}
\caption{An overview of the different necessary and sufficient conditions for MR which are discussed in this work for the different data sets, where $n$ is the number of measurement times (See accompanying text in Appendix \ref{app:mrzoo}). The different components of the data set are the expectation values of the operator $Q$ at times $t_i$ $\nexpval{Q_i}$, and the second, third, and fourth-order correlators, which are denoted as $C_{ij}$, $D_{ijk}$, and $E_{ijkl}$, respectively. In this work, the conditions explored are for a dichotomic variable in Sec.~\ref{sec:higher_correlators} and Sec.~\ref{sec:the_PIs}, and for a trichotomic variable in Sec.~\ref{sec:many_valued_variables}. {FTC denotes ``for the cycle".}}
\label{tbl:conditions}
\end{table*}

\begin{acknowledgments} 
We would like to thank Amanda Maity and Clement Mawby for useful discussions, and Andrew Cameron and Lane G. Gundermann for an initial reading of the manuscript. This work was funded in part by the Canadian federal government, the Ontario provincial government, and Mike and Ophelia Lazaridis.
\end{acknowledgments}

\appendix

\section{Organizing the conditions for MR}\label{app:mrzoo}

The number of different data sets and MR conditions explored in Sec.~\ref{sec:higher_correlators} and \ref{sec:the_PIs} may be unfamiliar to many readers so here, to put them in context, we describe the variety of possibilities for measurements made at up to five times. This is at the expense of some repetition of the main text but for clarity it is useful to have all the explanation of the conditions in the same place. We consider first conditions for a single dichotomic variable. What follows is depicted in Table \ref{tbl:conditions} which should be read in conjunction with the text.

At two times there is only one interesting possibility which is to measure the two averages $\langle\hQ_1 \rangle$, $\langle\hQ_2 \rangle$, and the correlator $C_{12}$. The MR conditions are the LG2s, Eq.~\eqref{eq:LG2}. In this case, the data set determines the underlying probability uniquely (it is proportional to the LG2s).

At three times, the standard approach is to determine the three averages $\langle\hQ_i \rangle$ and three correlators $C_{ij}$. The necessary and sufficient conditions for MR are the twelve LG2s and the four LG3s. These conditions ensure that that is some underlying three-time probability matching the data but does not determine it uniquely. This then leads to the second possibility which is to measure the third order correlator in addition to the standard data set. This determines the underlying probability uniquely, when it exists, and the third order condition Eq.~\eqref{3rdOrderLGIs} is precisely the condition that this probability is non-negative \citep{halliwell2019leggett}. In this case, the LG2s and LG3s do not need to be imposed separately since they are implied by the third order condition.

At four times, there are six possible two-time correlators. The standard data set consists of any cycle of four correlators $C_{ij}$, for example $C_{12}, C_{23}, C_{34}, C_{14}$, plus the four averages $\langle\hQ_i \rangle$ and the MR conditions consist of the eight LG4s for that cycle, together with {the cycle of LG2s that contain these correlators}. One could also consider a data set in which all six correlators are fixed. The MR conditions in this case turn out to be the set of all possible LG3s and LG2s \citep{halliwell2019fine} (Yet another option of course would be to fix five of the six correlators but the conditions in this case have not been worked out).

At four times, one also has the option of including third and fourth order correlators in the data set. Clearly there is quite a number of different choices of data sets here if only some of all the possible correlators are included. However, we focus only on the case in which all possible third and fourth order correlators are included and the MR conditions are then the fourth order LGIs, which fix the underlying four-time probability uniquely, where it exists \cite{halliwell2019leggett}. Again only the fourth order condition, Eq.~\eqref{4thOrderLGIs}, is needed here since it implies all lower order conditions. This general story concerning higher-order correlators is readily generalized to $n$ times, but we do not consider higher-order correlators beyond $n=4$.

At five times, there are ten two-time correlators. The standard data set sets of any cycle of five plus the five averages, and the MR conditions are the LG5s together with the LG2s for the chosen pairs of times. However, one could also consider fixing six, seven, eight, nine or ten two-time correlators. We focus in this paper on the case in which all ten are fixed. (The other cases have not been considered). The MR conditions are then the pentagon inequalities, together with all possible LG2s and LG3s \citep{halliwell2019fine}.

For the trichotomic case considered in Sec.~\ref{sec:many_valued_variables}, the MR conditions considered are the more familiar LG2s, LG3s and LG4s, but extended to the case of many-valued variables, which as discussed, involves employing more than one dichotomic variable at each time obeying familiar-looking LG inequalities.

{
\section{Interference terms and no signaling in time conditions} \label{app:nsit}

As indicated in the Introduction, from a quantum-mechanical perspective LG violations arise due to the presence of interference terms and we now describe this in more detail for the specific MR conditions discussed in this paper, following Refs.~\citep{halliwell2016leggett,halliwell2020conditions}. There is also a close relation between these interference terms and the no-signaling in time (NSIT) conditions \citep{kofler2013condition, clemente2015necessary}.

The NSIT condition at two times is the requirement that the quantity
\beq
I(s_2) = p_2 (s_2) - \sum_{s_1} p_{12} (s_1, s_2),
\eeq
is zero, where
\begin{equation}
    p_{12} (s_1, s_2) = \Tr(P_{s_2}(t_2) P_{s_1}(t_1) \rho P_{s_1}(t_1)) \label{eq:form_p12}
\end{equation}
denotes the probability for a pair of sequential measurements of $\hat Q$ at times $t_1, t_2$  and $p_2(s_2)$ denotes the probability of a single time measurement at $t_2$ with no earlier measurement. The NSIT condition is not satisfied when interferences are present so $I(s_2)$ is a measure of interference. It has the explicit form
\beq
I(s_2) = \frac{1}{4} \left( \langle \hat Q_2 \rangle - \langle \hat Q_2^{(1)} \rangle \right),
\eeq
where $ \langle \hat Q_2^{(1)} \rangle $ denotes the average of $\hat Q_2 $ in the presence of an earlier measurement at $t_1$ which has been summed out.
As show in Refs.~\citep{halliwell2016leggett, halliwell2020conditions}, the LG2 inequalities are violated when the interference term $I(s_2)$ is sufficiently large.

NSIT conditions have been used \cite{kofler2013condition, clemente2015necessary} to define a notion of macrorealism that is stronger than that considered here, in terms of LG inequalities. These different notions are compared in Refs.~\citep{halliwell2019necessary, halliwell2017comparing}. Here we use the NSIT conditions as more refined detectors of interferences compared to the LG inequalities which can be used to home in on particular interference terms. Note also that they involve different types of measurements of the same systems -- the NSIT conditions entail measurements in which everything is determined by a single experiment with sequential measurements but in the LG inequalities the results of a number of different experiments are combined.

At three times, a similar story arises. The interference terms that may cause violations of the LG3s are a linear combination of the terms $( C_{23} - C_{23}^{(1)}) $ and $ (C_{13} - C_{13}^{(2)}) $, where here and in what follows $C_{ij}^{(M)}$ denotes the correlator $C_{ij}$ in the presence of an intermediate or earlier measurement at time or times numbered by $M$. These interference terms may also be identified using three-time NSIT conditions, for example
\beq
I(s_1, s_3) = p_{13} (s_1,s_3) - \sum_{s_2} p_{123} (s_1,s_2,s_3),
\eeq
since it is easily seen that 
\beq
C_{13} - C_{13}^{(2)} = \sum_{s_1 s_3} \ s_1 s_3 I(s_1,s_3).
\eeq

Consider now the third order LGI, Eq.~\eqref{3rdOrderLGIs} in Sec.~\ref{sub2:conditions}. From the analysis in Ref.~\cite{halliwell2020conditions} it is readily seen that violations arise from a linear combination of the interference terms 
$ ( \langle \hat Q_2 \rangle - \langle \hat Q_2^{(1)} \rangle )$,
$( C_{23} - C_{23}^{(1)}) $, $ (C_{13} - C_{13}^{(2)})$ and 
 $( \langle \hat Q_3 \rangle - \langle \hat Q_3^{(12)} \rangle )$. The first three are familiar already from the LG2s and LG3s, but the final term is a new one in comparison, which may also be expressed in terms of a NSIT condition:
\beq
 \langle \hat Q_3 \rangle - \langle \hat Q_3^{(12)} \rangle  = \sum_{s_1 s_2 s_3} \ s_3
\left( p_{3} (s_3) - \sum_{s_1 s_2} p_{123} (s_1,s_2,s_3) \right).
\eeq
Hence the violation comes from two different types of interference term -- familiar ones arising in lower order LG inequalities plus a new one not seen at lower orders.

Turning now to the five-time models in Sec.~\ref{sec:the_PIs}, for the conditions involving the pentagon inequalities, there are ten correlators and nine interference terms. The first two are the same as those for the LG3s, above. There are three of the form $ (C_{14} - C_{14}^{(23)} )$ (plus two similar ones for $C_{24}$ and $C_{34}$) and four of the form $(C_{15} - C_{15}^{(234)})$ (plus three similar ones for $C_{25}, C_{35}, C_{45}$. The last four are clearly the most interesting since they cannot arise in any conditions at less than five times, so represent genuinely new features at five times. Again all these interference terms can be identified separately through NSIT conditions, as above.

For the models in Sec.~\ref{sec:many_valued_variables} involving many-valued variables, the analysis of interferences and NSIT conditions is similar to the LG3 case briefly summarized above, since the inequalities are phrased in terms of sets of dichtomic variables. However, there are also new possibilities arising from the fact that a dichotommic variable for a system of three or more levels can be measured in a number of different but macrorealistically equivalent ways which produces new types of interference terms \cite{pan2018quantum,budroni2014temporal,dakic2014density,wang2017enhanced, kumari2018violation}. This is all reviewed in Ref.~\cite{halliwell2020conditions}.}

\section{L\"{u}ders bound for higher-order LGIs}\label{app:luders}

Although LG inequalities of all the above types have algebraic lower bounds if the correlators could be chosen freely,
quantum theory typically imposes restrictions on how big the violations can be, which are usually greater than the algebraic bound. The two and three-time LG inequalities given above have a lower bound of $-1/2$ in quantum theory. This is the so-called L\"uders bound \citep{budroni2014temporal, dakic2014density, wang2017enhanced, kumari2018violation, pan2018quantum} and is readily derived as follows. Suppose for example we consider the LG2 for times $t_2$, $t_3$, we have
\begin{equation}
1 + \langle \hQ_2  \rangle + \langle \hQ_3 \rangle + C_{23} = \frac{1}{2} \langle ( 1 + \hQ_2 + \hQ_3 )^2 \rangle - \frac{1}{2},
\label{LG23jh}
\end{equation}
from which the lower bound is established, with equality if and only if,
\begin{equation}
( 1 + \hQ_2 + \hQ_3 ) | \psi \rangle = 0.
\label{con23}
\end{equation}
We now derive similar conditions for the third and fourth order LG inequalities given above.

The third order condition, Eq.~\eqref{3rdOrderLGIs}, may be written as $ \langle \hat K \rangle $, where the operator $\hat K$ is given by
\begin{align}
    \hat K =& \: 1 + \hQ_1 + \hQ_2 + \hQ_3 \nonumber \\
    &+ \frac{1}{2} \left(  \{ \hQ_1, \hQ_2 \} +
    \{ \hQ_2, \hQ_3 \}+ \{ \hQ_1, \hQ_3 \} \right)
    \nonumber \\
    &+ \frac{1}{4}  \{ \hQ_1, \{ \hQ_2, \hQ_3 \} \}.
\end{align}
With a small amount of algebra, this may be rewritten
\begin{equation}
    \hat K = {\hat P}_1 \hat B + \hat B P_1
\end{equation}
where ${\hat P}_1$ is the projector ${\hat P}_1 = (1 + \hat Q_1)/2$ and
\begin{equation}
\hat B = \frac{1}{2} (\hQ_2 + \hQ_3 + 1)^2 - \frac{1}{2}.
\end{equation}
Intuitively one might expect that $ \langle \hat K \rangle$ is most negative for a state $| \phi \rangle$ which satisfies
\begin{align}
    {\hat P}_1 | \phi \rangle &= | \phi \rangle, 
    \label{conP}\\
    \hat B | \phi \rangle &= \lambda | \phi \rangle,
    \label{conB}
\end{align}
where $\lambda $ is taken to be the most negative eigenvalue of $\hat B$, which can clearly be no smaller than $ -  \frac{1}{2}$. It follows then that 
\begin{equation}
\langle \hat K \rangle \ge -1.
\end{equation}
To prove this note that
\begin{equation}
\langle \hat K \rangle = \langle \psi| {\hat P}_1 \hat B | \psi \rangle +  \langle \psi| \hat B {\hat P}_1  | \psi \rangle,
\label{KBP}
\end{equation}
i.e.~it is twice the real part of the overlap between the vectors $ \hat B | \psi \rangle$ and $ {\hat P}_1 | \psi \rangle $. This will achieve its most negative value when the vectors satisfy
\begin{equation}
\hat B | \psi \rangle = - \alpha {\hat P}_1 | \psi \rangle, 
\label{conBP}
\end{equation}
for some $\alpha > 0$. (This is reasonably obvious but can be shown more decisively in a few lines using normalized vectors). It then follows that 
\begin{align}
    \langle \hat K \rangle &= -2 \alpha \langle \psi | {\hat P}_1 | \psi \rangle
    \nonumber \\
    &= 2 \langle \psi | \hat B | \psi \rangle.
\end{align}
This is bounded from below by $-1$, with equality when Eq.~\eqref{conB} holds and $\lambda = - \frac{1}{2}$, its smallest possible value. This in turn implies that Eq.~\eqref{conP} holds. This proves the result.

The extension to the fourth order case is readily accomplished from the observation that in the third order case we have
\begin{equation}
\hat K = 2 \{ {\hat P}_1, \{ {\hat P}_2, {\hat P}_3 \} \},
\label{K123}
\end{equation}
where ${\hat P}_i = (1 + \hat Q_i)/2 $,
and we have therefore shown that this operator has lowest eigenvalue $-1$.
The fourth order condition Eq.~\eqref{4thOrderLGIs} may be written in a similar form: $ \langle \hat K_{1234} \rangle \ge 0 $, where
\begin{equation}
\hat K_{1234} = 2 \{ {\hat P}_1, \{ {\hat P}_2,  \{ {\hat P}_3, {\hat P}_4 \} \} \},
\end{equation}
and note that this may be rewritten as
\begin{equation}
\hat K_{1234} = \hat P_1 \hat K_{234} + \hat K_{234} \hat P_1,
\end{equation}
where $\hat K_{234}$ is the operator Eq.(\ref{K123}) for the third order conditions at times $t_2, t_3,t_4$. We see that $\langle \hat K_{1234} \rangle $ then has exactly the same mathematical form as Eq.(\ref{KBP}), with $\hat B$, whose lowest eigenvalue is $- \frac{1}{2}$, replaced by $\hat K_{234}$, whose lowest eigenvalue is $ -1$. It follows from an identical argument that
\begin{equation}
\langle \hat K_{1234} \rangle \ge -2
\end{equation}
with equality when $ \hat P_1 | \psi \rangle = | \psi \rangle $ and when $ \hat K_{234} |\psi \rangle = - | \psi \rangle $ (which in turn holds when the conditions analogous to Eqs.(\ref{conB}), (\ref{conP}) hold). This established the L\"uders bound for the fourth order case.
The method of proof clearly generalizes to higher order cases.

\section{Correlators for projective measurements}\label{app:corr_forms}

The second-order correlator, is defined classically as 
\begin{equation}
    C_{ij} = \nexpval{Q_iQ_j}.
\end{equation}
In the quantum case, $C_{ij}$ is determined from the two-time measurement probability {$p_{ij}(s_i,s_j)$, Eq.~\eqref{eq:form_p12}, }by \citep{fritz2010quantum, majidy2019violation}
\begin{equation}
    C_{ij} = \sum_{s_i, s_j} s_i s_j p_{ij}(s_i,s_j),
\end{equation}
which is readily shown to yield 
\begin{align}
    C_{ij} =& \frac{1}{2}\nexpval{\hQ_j\hQ_i + \hQ_i\hQ_j}.
\end{align}
We now follow the same derivation used in Ref.~\citep{fritz2010quantum} to find the form $C_{ij}$, to determine the form of $D_{ijk}$ and $E_{ijkl}$ for a quantum system measured via projective measurements. 

We first define $P_{s_i}(t_i) \equiv (1+s_i\hQ(t_i))/2$ as the projector onto the $s_i$ eigenstate of $\hQ_i$. In this derivation, and the following ones, we will make repeated use of the following two results,
\begin{align}
    \sum_{s_i} s_i P_{s_i}(t_i)\rho P_{s_i}(t_i) &= \frac{1}{2} (\hQ_i\rho + \rho \hQ_i)\\
    \sum_{s_i} s_i P_{s_i}(t_i)\rho &= \hQ_i\rho.
\end{align}
We find then that,
\begin{align}
    D_{ijk} =& \sum_{s_i, s_j, s_k} s_i s_j s_k p(s_i, s_j, s_k)\\
    =& \sum_{s_i, s_j, s_k} s_i s_j s_k \text{tr} [P_{s_k}(t_k)P_{s_j}(t_j) P_{s_i}(t_i)\rho  \nonumber \\&P_{s_i}(t_i) P_{s_j}(t_j)] \\
    =& \frac{1}{4} \text{tr} [\hQ_{k}\hQ_{j}\hQ_{i}\rho + \hQ_{k}\hQ_{j}\rho \hQ_{i} + \hQ_{k}\hQ_{i}\rho \hQ_{j}  \nonumber \\ & + \hQ_{k}\rho \hQ_{i} \hQ_{j})] \\
    =& \frac{1}{4}\nexpval{\acomm{\hat{Q}_i}{\acomm{\hat{Q}_j}{\hat{Q}_k}}}.
\end{align}
{Note that this form for $D_{ijk}$ was found earlier in Ref.~\citep{berg2009commuting}, using a different approach.} Similar to how the form of $D_{ijk}$ follows from $C_{ij}$, by applying an additional projection, the form of $E_{ijkl}$ can readily be found to be,
\begin{equation}
    E_{ijkl} =  \frac{1}{8}\nexpval{\acomm{\hat{Q}_i}{\acomm{\hat{Q}_j}{\acomm{\hat{Q}_k}{\hat{Q}_l}}}}, 
\end{equation}

\section{Models for studying spin-1 systems}\label{app:simple_spin_1_model}

In this Appendix we discuss two models for studying a spin-1 system. The first model, which is used in Sec.~\ref{sec:higher_correlators}, considers a dichotomic operator. The other model, which is used in Sec.~\ref{sec:many_valued_variables}, consider a trichotomic operator. In both cases the systems Hamiltonian is $\hat{H} = \omega \htsg_x/2$, where $\htsg_x$ is the three-level Pauli-x matrix,
\begin{equation}
    \hat{H} = \frac{\omega}{2\sqrt{2}}
    \begin{pmatrix}
    0 & 1 & 0\\
    1 & 0 & 1 \\
    0 & 1 &  0 
    \end{pmatrix}.
\end{equation}
As well, the general initial state for the three-level system is parameterized as \citep{mendavs2006classification}, 
\begin{equation}
    \rho = \begin{pmatrix}
    \frac{1}{3} + a_3 + \frac{a_8}{\sqrt{3}} & a_1 - ia_2 & a_4 - ia_5 \\
    a_1 + ia_2 & \frac{1}{3} - a_3 + \frac{a_8}{\sqrt{3}} & a_6 - ia_7 \\
     a_4 + ia_5 &  a_6 + ia_7 & \frac{1}{3} - \frac{2a_8}{\sqrt{3}}
    \end{pmatrix}, \label{eq:general_3_level_rho}
\end{equation}
where $a_i$ are real parameters, whose values are bounded by the conditions $\mathbf{a} \cdot \mathbf{a} \leq \frac{1}{3}$ and $ 0 \leq \det(\rho)$, where $\mathbf{a} \equiv (a_1, a_2, \hdots, a_8)$.

\subsection{Model for a dichotomic operator}\label{app:spin_1_dichotomic}

In this model we choose our dichotomic operator to be,
 \begin{equation}
    \hQ =  \begin{pmatrix}
    -1 & 0 & 0\\
    0 & 1 & 0\\
    0 & 0 & -1
    \end{pmatrix} \label{eq:Q_dichotomic}.
\end{equation}
It is easy to check that $\hQ$ anti-commutes with $\hat{H}$ and satisfies the desired properties for a dichotomic measurement operator (that it is a Hermitian operator with $\pm 1$ eigenvalues). For the spin-1 Hamiltonian we have that the time evolution is,
\begin{align}
    e^{- i\omega t_i\sigma_x/2} &= 1 - i\htsg_x \sti + \htsg_x^2(\cti - 1).
\end{align}
Using the anti-commutation relationship between $\hQ$ and $\htsg_x$ we find that $\hQ_i$ and its products can be written as,
\begin{align}
    \hQ_i &= \hQ + i\sin(\omega t_i)\htsg_x\hQ \nonumber \\ 
    &+ (\cos(\omega t_i) - 1) \htsg_x^2\hQ\\
     \hQ_i\hQ_j &= I + i\sin(\omega t_i -\omega t_j )\htsg_x  \nonumber \\ 
    & + (\cos(\omega t_i -\omega t_j) \smin 1)\htsg_x^2\\
    \hQ_i \hQ_j \hQ_k =& \: \hQ + i\sin(\omega t_i - \omega t_j + \omega t_k)\htsg_x \hQ  \nonumber \\& + (\cos(\omega t_i - \omega t_j + \omega t_k)-1)\htsg_x^2 \hQ\\
    \hQ_i \hQ_j \hQ_k \hQ_l =& \: \one + i\sin(\omega t_i - \omega t_j + \omega t_k - \omega t_l)\htsg_x   \nonumber \\& + (\cos(\omega t_i - \omega t_j + \omega t_k - \omega t_l)-1)\htsg_x^2.
\end{align}
Using these results we find that
\begin{align}
    C_{ij} =& \frac{1}{2} \nexpval{\acomm{Q_i}{Q_j}} \nonumber\\=& 1 + (\cos(\omega t_i - \omega t_j) \smin 1)\nexpval{\htsg_x^2}\\
    D_{ijk} =& \frac{1}{4}\nexpval{\acomm{\hat{Q}_i}{\acomm{\hat{Q}_j}{\hat{Q}_k}}} \\
    =& \nexpval{\hQ} + i\sin(\omega t_i)\cos(\omega t_j - t_k) \nexpval{\htsg_x\hQ} \nonumber\\&
    + \big(\cos(\omega t_i)\cos(\omega t_j - \omega t_k) - 1 \big) \nexpval{\htsg_x^2\hQ}\\
    E_{ijkl} =& \frac{1}{8}\nexpval{\acomm{\hat{Q}_i}{\acomm{\hat{Q}_j}{\acomm{\hat{Q}_k}{\hat{Q}_l}}}},\\
    =& 1
    + \big(\cos(\omega t_i - \omega t_j)\cos(\omega t_k - \omega t_l)  - 1 \big)\nexpval{\htsg_x^2\hQ}.
\end{align}
Thus we have all of the correlators necessary for our investigation in Sec.~\ref{sec:higher_correlators}. 

\subsection{Model for a trichotomic operator}\label{app:gen_op_form}

In Sec.~\ref{sec:many_valued_variables}, we consider a simple spin-1 model measured with the operators $\hQ$, $\hR$, and $\hS$, constructed from the states $\ket{A}$, $\ket{B}$, and $\ket{C}$. We have found that a simple and convenient choice for these states is
\begin{align}
    \ket{A} &= \frac{1}{\sqrt{2}}(\ket{\E_+} - \ket{\E_-})\\
  \ket{B} &= \frac{1}{\sqrt{2}}(\ket{\E_+} + \ket{\E_-})\\
    \ket{C} &= \ket{\E_0},
\end{align}
for which the operators are:
\begin{align}
    \hQ &=  - \dyad{\E_0}{\E_0} - \dyad{\E_+}{\E_-} - \dyad{\E_-}{\E_+}\\
    \hR &= - \dyad{\E_0}{\E_0} + \dyad{\E_+}{\E_-} + \dyad{\E_-}{\E_+}\\
    \hS &= \dyad{\E_0}{\E_0} - \dyad{\E_+}{\E_+} - \dyad{\E_-}{\E_-}.
\end{align}
Here a factor of $-1$ is applied to each operator, so that each has two $-1$ eigenvalues.

These operators have the convenient properties that $\hQ$ and $\hR$ anti-commute with $\hat{H}$ and $\hS$ commutes with $\hat{H}$. The resulting model has some resemblance to the simple spin-1/2 model, in that the dynamics primarily concerns transitions between just two states.

Due to the fact that $\hQ+\hR+\hS=-1$, all averages and correlators can be computed from the following:
\begin{align}
 \langle \hQ_i \rangle =& \expval{Q} + i\sin(\omega t_i)\expval{\htsg_xQ}   \nonumber \\ &+ (\cos(\omega t_i) - 1) \expval{\htsg_x^2Q}\\
  \langle \hR_i \rangle =& \expval{R} - i\sin(\omega t_i)\expval{\htsg_xQ} \nonumber \\ & - (\cos(\omega t_i) - 1) \expval{\htsg_x^2Q}\\
  \langle\hQ_i\hQ_j \rangle =& \langle\hR_i\hR_j \rangle =  1 + (\cos(\omega_i  -  \omega_j)  -  1)\expval{\htsg_x^2}\\
  \langle\hQ_i\hR_j \rangle =& \langle\hR_i\hQ_j \rangle =\expval{S} - (\cos(\omega_i  -  \omega_j)  -  1) \expval{\htsg_x^2}.
\end{align}
Note that $\langle \hS \rangle = 1 - 2\htsg_x^2$. The relevant LG inequalities for this trichotomic system are given by Eq.~\eqref{eq:3lvlLG2} and \eqref{eq:3lvlLG3} and are also written out in full in the appendix of Ref.~\citep{halliwell2020conditions}.

\section{Parameterization for Sec.~III} \label{sec:Neq4b}

\begin{table}
\centering
\begin{tabular}{@{}ccccc@{}}
\toprule
Case & Level & $\hat{Q}$ & $\ket{A}$ & $\ket{B}$ \\ \midrule
1 & \multirow{2}{*}{2} & \multirow{2}{*}{$1 - 2 \dyad{A}$} & $\ket{1}$ & -- \\
2 &  &  & $\ket{2}$ & -- \\ \midrule
3 & \multirow{3}{*}{3} & \multirow{3}{*}{$1 - 2 \dyad{A}$} & $\ket{1}$ & -- \\
4 &  &  & $\ket{2}$ & -- \\
5 &  &  & $\ket{3}$ & -- \\ \midrule
6 & \multirow{4}{*}{4} & \multirow{4}{*}{$1 - 2 \dyad{A}$} & $\ket{1}$ & -- \\
7 &  &  & $\ket{2}$ & -- \\
8 &  &  & $\ket{3}$ & -- \\
9 &  &  & $\ket{4}$ & -- \\ \midrule
10 & \multirow{6}{*}{4} & \multirow{6}{*}{$1 - 2 \dyad{A} - 2\dyad{B}$} & $\ket{1}$ & $\ket{2}$ \\
11 &  &  & $\ket{1}$ & $\ket{3}$ \\
12 &  &  & $\ket{1}$ & $\ket{4}$ \\
13 &  &  & $\ket{2}$ & $\ket{3}$ \\
14 &  &  & $\ket{2}$ & $\ket{4}$ \\
15 &  &  & $\ket{3}$ & $\ket{4}$ \\ \bottomrule
\end{tabular}
\caption{The fifteen cases necessary to investigate all form of $\hat{Q}$ for a two- to four-level system. Here $\ket{A}$ and $\ket{B}$ are the general labels introduced in the text and $\ket{1},\ket{2},\ket{3},$ and $\ket{4}$ are the labels for four basis states in the $N$ dimensional $\htsg_z$ basis. The list of cases is derived in Sec.~\ref{sec:the_PIs}.}
\label{tab:15_cases}
\end{table}

\begin{table*}
\resizebox{\textwidth}{!}{%
\begin{tabular}{@{}clll@{}}
\toprule
N & \multicolumn{1}{c}{$\ket{v_i}^{(n)}$} & \multicolumn{1}{c}{$\ip{v_i}{v_j}$} & \multicolumn{1}{c}{$\ip{\psi}{\E}$} \\ \midrule  \vspace{3pt}
2  & \rule{0pt}{21pt}{$\!\begin{aligned}  
    \ket{v_i}^{(1)} &= \tfrac{1}{\sqrt{2}} \big( e^{-i\omega t_i/2} \ket{\E_+} +  e^{i\omega t_i/2} \ket{\E_-} \big)\\
    \ket{v_i}^{(2)} &= \tfrac{1}{\sqrt{2}}  \big(e^{-i\omega t_i/2} \ket{\E_+} - e^{i\omega t_i/2} \ket{\E_-}  \big)
    \end{aligned}$}
& {$\!\begin{aligned} 
    \ip{v_j}{v_i}^{(1)} &= \ip{v_j}{v_i}^{(2)}\\
    &= \cos(\tfrac{\omega}{2}(t_j - t_i))
 \end{aligned}$}
& {$\!\begin{aligned} 
 \langle \psi | \E_+ \rangle &= \cos\theta\\
    \langle \psi | \E_- \rangle &= e^{i\phi} \sin \theta
 \end{aligned}$}
\\ \hline \vspace{3pt}
3 & \rule{0pt}{31pt}{$\!\begin{aligned}  
    \ket{v_i}^{(3)} &= \tfrac{1}{2} \big( e^{-i\omega t_i}\ket{\E_+} +  \sqrt{2}\ket{\E_0} + e^{i\omega t_i}\ket{\E_-} \big) \\
    \ket{v_i}^{(4)} &= \tfrac{1}{\sqrt{2}} \big( e^{-i\omega t_i}\ket{\E_+} -  e^{i\omega t_i}\ket{\E_-} \big) \\
    \ket{v_i}^{(5)} &= \tfrac{1}{2} \big( e^{-i\omega t_i}\ket{\E_+} -  \sqrt{2}\ket{\E_0} + e^{i\omega t_i}\ket{\E_-} \big)
    \end{aligned}$}
& {$\!\begin{aligned} 
    \ip{v_j}{v_i}^{(3)} &= \ip{v_j}{v_i}^{(5)} \\
    &= \tfrac{1}{2} \left( 1 + \cos \left( \omega t_j - \omega t_i \right) \right)\\
    \ip{v_j}{v_i}^{(4)} &= \cos(\omega t_j - \omega t_i)
 \end{aligned}$}
& {$\!\begin{aligned} 
 \langle \psi | \E_+ \rangle &= \cos \theta \\
    \langle \psi | \E_0 \rangle &= e^{ i \phi_1} \sin \theta \cos \alpha\\
    \langle \psi | \E_- \rangle &= e^{i \phi_2} \sin \theta \sin \alpha
 \end{aligned}$}
\\ \hline  \vspace{3pt}
4
& \rule{0pt}{75pt}{$\!\begin{aligned}
    \ket{v_i}^{(6)} =&\tfrac{-1}{2\sqrt{2}}\big(e^{i3\omega t_i/2} \ket{\E_{-3}} - \sqrt{3} e^{i\omega t_i/2} \ket{\E_{-1}} \nonumber \\
    &+ \sqrt{3} e^{-i\omega t_i/2} \ket{\E_{+1}} - e^{-i3\omega t_i/2} \ket{\E_{+3}}\big) \\
    \ket{v_i}^{(7)}
    =&\tfrac{1}{2\sqrt{2}}\big( \sqrt{3} e^{i3\omega t_i/2} \ket{\E_{-3}} - e^{i\omega t_i/2} \ket{\E_{-1}} \nonumber \\
    &- e^{-i\omega t_i/2} \ket{\E_{+1}} + \sqrt{3}e^{-i3\omega t_i/2} \ket{\E_{+3}}\big)\\
    \ket{v_i}^{(8)} =&\tfrac{-1}{2\sqrt{2}}\big(\sqrt{3}e^{i3\omega t_i/2} \ket{\E_{-3}} + e^{i\omega t_i/2} \ket{\E_{-1}} \nonumber \\
    &- e^{-i\omega t_i/2} \ket{\E_{+1}} - \sqrt{3} e^{-i3\omega t_i/2} \ket{\E_{+3}}\big)\\
    \ket{v_i}^{(9)} =&\tfrac{1}{2\sqrt{2}}\big(e^{i3\omega t_i/2} \ket{\E_{-3}} + \sqrt{3} e^{i\omega t_i/2} \ket{\E_{-1}} \nonumber \\
    &+ \sqrt{3} e^{-i\omega t_i/2} \ket{\E_{+1}} + e^{-i3\omega t_i/2} \ket{\E_{+3}}\big)
    \end{aligned}$}
& {$\!\begin{aligned} 
    \ip{v_j}{v_i}^{(6)} &= \ip{v_j}{v_i}^{(9)}\\ 
    &= \tfrac{1}{4} \cos(\tfrac{3\omega}{2} (t_j - t_i)) \\& +  \tfrac{3}{4} \cos(\tfrac{\omega}{2} (t_j - t_i)) \\
    \ip{v_j}{v_i}^{(7)} &= \ip{v_j}{v_i}^{(8)}  \\
    &= \tfrac{3}{4} \cos(\tfrac{3\omega}{2} (t_j - t_i)) \\& +  \tfrac{1}{4} \cos(\tfrac{\omega}{2} (t_j - t_i))
 \end{aligned}$}
& {$\!\begin{aligned} 
    \langle \psi | \E_{+3} \rangle &=  \cos \theta\\
    \langle \psi | \E_{+1} \rangle &=  e^{i \phi_1} \sin \theta \cos \alpha\\
    \langle \psi | \E_{-1}\rangle &= e^{ i \phi_2} \sin \theta \sin \alpha \cos \beta \\
    \langle \psi | \E_{-3} \rangle &= e^{i \phi_3} \sin \theta \sin \alpha \sin \beta \\
 \end{aligned}$}
\\ \bottomrule
\end{tabular}
}
\caption{The values of $\ket{v_i}^{(n)}$, $\ip{v_i}{v_j}$, and the parameterization used for $\ip{\psi}{\E}$ for each case in Table \ref{tab:15_cases}.}
\label{tab:all_vi}
\end{table*}

We outline here the derivation of $\ket{v_i}$ and $\ip{v_j}{v_i}$, and the parameterization of $\ket{\psi}$ for each of the different cases in which a numerical search was done for in Sec.~\ref{sec:the_PIs}. There are fifteen cases in total for an $N=2$ to $4$ level system and are given in Table \ref{tab:15_cases}.

For each case, our procedure is as follows. We identify the Hamiltonian $\hat{H}$ and find its eigenstates $\ket{\E_n}$. Depending on the cases, the projection of the eigenstate onto $\ket{A}$ will be different. Here we choose the choices of $\ket{A}$ to be the basis states in the $\htsg_z$ basis. We then find $\ip{\E_n}{A}$, which we use with Eq.\eqref{eq:visum} to find $\ket{v_i}^{(N)}$, where the superscript in parentheses is used to distinguish between the different cases. Using this vector the form of $\ip{v_i}{v_j}$ readily follows. What remains then is to parameterize $\ip{\psi}{\E_n}$ such that the initial state is normalized. 

$\ket{v_i}$ and $\ip{v_j}{v_i}$, and the parameterization for $\ket{\psi}$ for the first 9 of the 15 cases are outlined in Table \ref{tab:all_vi}. The Hamiltonians in each case is the $N$ dimensional spin Hamiltonian \citep{levitt2013spin}. The eigenstates references in the table are - For the 2 level system:
\begin{equation}
    \ket{\E_{\pm}} = \frac{1}{\sqrt{2}} \begin{pmatrix} 1 \\ \pm 1 \end{pmatrix},
\end{equation}
with eigenvalues $\E_{\pm} = \pm \frac{1}{2}$, for the three-level system:
\begin{equation}
    | \E_{\pm} \rangle = \frac{1}{2} \begin{pmatrix} 1 \\ \pm \sqrt{2} \\ 1 \end{pmatrix}, \ \ \ \ 
    |\E_0 \rangle = \frac{1}{ \sqrt{2} }\begin{pmatrix} 1 \\ 0 \\ -1 \end{pmatrix}, \nonumber
\end{equation}
with eigenvalues $\E_{\pm} = \pm 1$ and $\E_{\pm} = \pm 0$, and for the four-level system:
\begin{align*}
    \ket{\E_{\pm3}} &=  \frac{1}{2\sqrt{2}}
    \begin{pmatrix} \pm 1 \\ \sqrt{3} \\ \pm \sqrt{3} \\ 1 \end{pmatrix}, \ \
    \ket{\E_{\pm1}} = \frac{1}{2\sqrt{2}}
    \begin{pmatrix} \mp \sqrt{3} \\ -1 \\ \pm  1 \\ \sqrt{3} \end{pmatrix}.
\end{align*}
with eigenvalues $\E_{\pm 3} = \pm \frac{3}{2}$ and $\E_{\pm 1} = \pm \frac{1}{2}$.

Cases 10 to 15 are formulated with a different operator,
\begin{equation}
    \hat Q_i ^{(4b)} = 1 - 2 \dyad{v_i} - 2\dyad{u_i}.
\end{equation}
For this operator the correlators take the form,
\begin{align}
    \langle \hat Q_i \rangle &= 1 - 2 \abs{\ip{\psi}{v_i}}^2 - 2 \abs{\ip{\psi}{u_i}}^2, \label{eq:general_Qib}\\
    C_{ij} 
    &= 1 - 2 \abs{\ip{\psi}{v_i}}^2  - 2 \abs{\ip{\psi}{v_j}}^2  - 2 \abs{\ip{\psi}{u_i}}^2 \nonumber \\
    & - 2 \abs{\ip{\psi}{u_j}}^2 
    + 4\Re(\ip{\psi}{v_i}\ip{v_i}{v_j}\ip{v_j}{\psi})\nonumber \\
    &+ 4\Re(\ip{\psi}{v_i}\ip{v_i}{u_j}\ip{u_j}{\psi})\nonumber \\
    &+ 4\Re(\ip{\psi}{u_i}\ip{u_i}{v_j}\ip{v_j}{\psi})\nonumber \\
    &+ 4\Re(\ip{\psi}{u_i}\ip{u_i}{u_j}\ip{u_j}{\psi})
    .\label{eq:general_Cijb}
\end{align}
The six cases correspond to the vectors $\ket{v_i}$ and $\ket{u_i}$ taking any combination of two of the four listed for $N=4$ in Table \ref{tab:all_vi}. Each search will require two sets of the parameterizations given in Table \ref{tab:all_vi}. We also need to consider the inner products of different $\ket{v_i}^{n}$, which we denote as $ \ip{v_j}{u_i}^{(n/m)}$, and are,
\begin{align}
    \ip{v_j}{u_i}^{(6/8)} =& \ip{v_j}{u_i}^{(7/9)} = \frac{\sqrt{3}}{4}(\cos(\frac{3\omega}{2}(t_j-t_i)) \nonumber \\& - \cos(\frac{\omega}{2}(t_j-t_i)))\\
    \ip{v_j}{u_i}^{(6/7)} =& \ip{v_j}{u_i}^{(8/9)} = \frac{-i\sqrt{3}}{4}(\sin(\frac{\omega}{2}(t_j-t_i)) \nonumber \\& + \sin(\frac{3\omega}{2}(t_j-t_i)))\\
    \ip{v_j}{u_i}^{(6/9)} =& \frac{i}{4}(3\sin(\frac{\omega}{2}(t_j-t_i)) - \sin(\frac{3\omega}{2}(t_j-t_i)))\\
    \ip{v_j}{u_i}^{(7/8)} =& \frac{i}{4}(\sin(\frac{\omega}{2}(t_j-t_i)) - 3\sin(\frac{3\omega}{2}(t_j-t_i))).
\end{align}

\vspace{1.0cm}

\nocite{*}
\bibliography{bib}

\providecommand{\noopsort}[1]{}\providecommand{\singleletter}[1]{#1}%
\begin{thebibliography}{60}%
\makeatletter
\providecommand \@ifxundefined [1]{%
 \@ifx{#1\undefined}
}%
\providecommand \@ifnum [1]{%
 \ifnum #1\expandafter \@firstoftwo
 \else \expandafter \@secondoftwo
 \fi
}%
\providecommand \@ifx [1]{%
 \ifx #1\expandafter \@firstoftwo
 \else \expandafter \@secondoftwo
 \fi
}%
\providecommand \natexlab [1]{#1}%
\providecommand \enquote  [1]{``#1''}%
\providecommand \bibnamefont  [1]{#1}%
\providecommand \bibfnamefont [1]{#1}%
\providecommand \citenamefont [1]{#1}%
\providecommand \href@noop [0]{\@secondoftwo}%
\providecommand \href [0]{\begingroup \@sanitize@url \@href}%
\providecommand \@href[1]{\@@startlink{#1}\@@href}%
\providecommand \@@href[1]{\endgroup#1\@@endlink}%
\providecommand \@sanitize@url [0]{\catcode `\\12\catcode `\$12\catcode
  `\&12\catcode `\#12\catcode `\^12\catcode `\_12\catcode `\%12\relax}%
\providecommand \@@startlink[1]{}%
\providecommand \@@endlink[0]{}%
\providecommand \url  [0]{\begingroup\@sanitize@url \@url }%
\providecommand \@url [1]{\endgroup\@href {#1}{\urlprefix }}%
\providecommand \urlprefix  [0]{URL }%
\providecommand \Eprint [0]{\href }%
\providecommand \doibase [0]{https://doi.org/}%
\providecommand \selectlanguage [0]{\@gobble}%
\providecommand \bibinfo  [0]{\@secondoftwo}%
\providecommand \bibfield  [0]{\@secondoftwo}%
\providecommand \translation [1]{[#1]}%
\providecommand \BibitemOpen [0]{}%
\providecommand \bibitemStop [0]{}%
\providecommand \bibitemNoStop [0]{.\EOS\space}%
\providecommand \EOS [0]{\spacefactor3000\relax}%
\providecommand \BibitemShut  [1]{\csname bibitem#1\endcsname}%
\let\auto@bib@innerbib\@empty
\bibitem [{\citenamefont {Friedman}\ \emph {et~al.}(2000)\citenamefont
  {Friedman}, \citenamefont {Patel}, \citenamefont {Chen}, \citenamefont
  {Tolpygo},\ and\ \citenamefont {Lukens}}]{friedman2000quantum}%
  \BibitemOpen
  \bibfield  {author} {\bibinfo {author} {\bibfnamefont {J.~R.}\ \bibnamefont
  {Friedman}}, \bibinfo {author} {\bibfnamefont {V.}~\bibnamefont {Patel}},
  \bibinfo {author} {\bibfnamefont {W.}~\bibnamefont {Chen}}, \bibinfo {author}
  {\bibfnamefont {S.}~\bibnamefont {Tolpygo}},\ and\ \bibinfo {author}
  {\bibfnamefont {J.~E.}\ \bibnamefont {Lukens}},\ }\href@noop {} {\bibfield
  {journal} {\bibinfo  {journal} {Nature}\ }\textbf {\bibinfo {volume} {406}},\
  \bibinfo {pages} {43} (\bibinfo {year} {2000})}\BibitemShut {NoStop}%
\bibitem [{\citenamefont {Ladd}\ \emph {et~al.}(2010)\citenamefont {Ladd},
  \citenamefont {Jelezko}, \citenamefont {Laflamme}, \citenamefont {Nakamura},
  \citenamefont {Monroe},\ and\ \citenamefont {O’Brien}}]{ladd2010quantum}%
  \BibitemOpen
  \bibfield  {author} {\bibinfo {author} {\bibfnamefont {T.~D.}\ \bibnamefont
  {Ladd}}, \bibinfo {author} {\bibfnamefont {F.}~\bibnamefont {Jelezko}},
  \bibinfo {author} {\bibfnamefont {R.}~\bibnamefont {Laflamme}}, \bibinfo
  {author} {\bibfnamefont {Y.}~\bibnamefont {Nakamura}}, \bibinfo {author}
  {\bibfnamefont {C.}~\bibnamefont {Monroe}},\ and\ \bibinfo {author}
  {\bibfnamefont {J.~L.}\ \bibnamefont {O’Brien}},\ }\href@noop {} {\bibfield
   {journal} {\bibinfo  {journal} {Nature}\ }\textbf {\bibinfo {volume}
  {464}},\ \bibinfo {pages} {45} (\bibinfo {year} {2010})}\BibitemShut
  {NoStop}%
\bibitem [{\citenamefont {Nielsen}\ and\ \citenamefont
  {Chuang}(2002)}]{nielsen2002quantum}%
  \BibitemOpen
  \bibfield  {author} {\bibinfo {author} {\bibfnamefont {M.~A.}\ \bibnamefont
  {Nielsen}}\ and\ \bibinfo {author} {\bibfnamefont {I.}~\bibnamefont
  {Chuang}},\ }\href@noop {} {\bibinfo {title} {Quantum computation and quantum
  information}} (\bibinfo {year} {2002})\BibitemShut {NoStop}%
\bibitem [{\citenamefont {Bell}(1964)}]{bell1964einstein}%
  \BibitemOpen
  \bibfield  {author} {\bibinfo {author} {\bibfnamefont {J.~S.}\ \bibnamefont
  {Bell}},\ }\href@noop {} {\bibfield  {journal} {\bibinfo  {journal} {Physics
  Physique Fizika}\ }\textbf {\bibinfo {volume} {1}},\ \bibinfo {pages} {195}
  (\bibinfo {year} {1964})}\BibitemShut {NoStop}%
\bibitem [{\citenamefont {Salart}\ \emph {et~al.}(2008)\citenamefont {Salart},
  \citenamefont {Baas}, \citenamefont {Branciard}, \citenamefont {Gisin},\ and\
  \citenamefont {Zbinden}}]{salart2008testing}%
  \BibitemOpen
  \bibfield  {author} {\bibinfo {author} {\bibfnamefont {D.}~\bibnamefont
  {Salart}}, \bibinfo {author} {\bibfnamefont {A.}~\bibnamefont {Baas}},
  \bibinfo {author} {\bibfnamefont {C.}~\bibnamefont {Branciard}}, \bibinfo
  {author} {\bibfnamefont {N.}~\bibnamefont {Gisin}},\ and\ \bibinfo {author}
  {\bibfnamefont {H.}~\bibnamefont {Zbinden}},\ }\href@noop {} {\bibfield
  {journal} {\bibinfo  {journal} {Nature}\ }\textbf {\bibinfo {volume} {454}},\
  \bibinfo {pages} {861} (\bibinfo {year} {2008})}\BibitemShut {NoStop}%
\bibitem [{\citenamefont {Palacios-Laloy}\ \emph {et~al.}(2010)\citenamefont
  {Palacios-Laloy}, \citenamefont {Mallet}, \citenamefont {Nguyen},
  \citenamefont {Bertet}, \citenamefont {Vion}, \citenamefont {Esteve},\ and\
  \citenamefont {Korotkov}}]{palacios2010experimental}%
  \BibitemOpen
  \bibfield  {author} {\bibinfo {author} {\bibfnamefont {A.}~\bibnamefont
  {Palacios-Laloy}}, \bibinfo {author} {\bibfnamefont {F.}~\bibnamefont
  {Mallet}}, \bibinfo {author} {\bibfnamefont {F.}~\bibnamefont {Nguyen}},
  \bibinfo {author} {\bibfnamefont {P.}~\bibnamefont {Bertet}}, \bibinfo
  {author} {\bibfnamefont {D.}~\bibnamefont {Vion}}, \bibinfo {author}
  {\bibfnamefont {D.}~\bibnamefont {Esteve}},\ and\ \bibinfo {author}
  {\bibfnamefont {A.~N.}\ \bibnamefont {Korotkov}},\ }\href@noop {} {\bibfield
  {journal} {\bibinfo  {journal} {Nature Physics}\ }\textbf {\bibinfo {volume}
  {6}},\ \bibinfo {pages} {442} (\bibinfo {year} {2010})}\BibitemShut {NoStop}%
\bibitem [{\citenamefont {Leggett}\ and\ \citenamefont
  {Garg}(1985)}]{leggett1985quantum}%
  \BibitemOpen
  \bibfield  {author} {\bibinfo {author} {\bibfnamefont {A.~J.}\ \bibnamefont
  {Leggett}}\ and\ \bibinfo {author} {\bibfnamefont {A.}~\bibnamefont {Garg}},\
  }\href@noop {} {\bibfield  {journal} {\bibinfo  {journal} {Physical Review
  Letters}\ }\textbf {\bibinfo {volume} {54}},\ \bibinfo {pages} {857}
  (\bibinfo {year} {1985})}\BibitemShut {NoStop}%
\bibitem [{\citenamefont {Leggett}(2008)}]{leggett2008realism}%
  \BibitemOpen
  \bibfield  {author} {\bibinfo {author} {\bibfnamefont {A.~J.}\ \bibnamefont
  {Leggett}},\ }\href@noop {} {\bibfield  {journal} {\bibinfo  {journal}
  {Reports on Progress in Physics}\ }\textbf {\bibinfo {volume} {71}},\
  \bibinfo {pages} {022001} (\bibinfo {year} {2008})}\BibitemShut {NoStop}%
\bibitem [{\citenamefont {Timpson}\ and\ \citenamefont
  {Maroney}(2013)}]{timpson2013quantum}%
  \BibitemOpen
  \bibfield  {author} {\bibinfo {author} {\bibfnamefont {C.}~\bibnamefont
  {Timpson}}\ and\ \bibinfo {author} {\bibfnamefont {O.}~\bibnamefont
  {Maroney}},\ }\href@noop {} {\bibfield  {journal} {\bibinfo  {journal} {The
  British Journal for the Philosophy of Science}\ } (\bibinfo {year}
  {2013})}\BibitemShut {NoStop}%
\bibitem [{\citenamefont {Emary}\ \emph {et~al.}(2013)\citenamefont {Emary},
  \citenamefont {Lambert},\ and\ \citenamefont {Nori}}]{emary2013leggett}%
  \BibitemOpen
  \bibfield  {author} {\bibinfo {author} {\bibfnamefont {C.}~\bibnamefont
  {Emary}}, \bibinfo {author} {\bibfnamefont {N.}~\bibnamefont {Lambert}},\
  and\ \bibinfo {author} {\bibfnamefont {F.}~\bibnamefont {Nori}},\ }\href@noop
  {} {\bibfield  {journal} {\bibinfo  {journal} {Reports on Progress in
  Physics}\ }\textbf {\bibinfo {volume} {77}},\ \bibinfo {pages} {016001}
  (\bibinfo {year} {2013})}\BibitemShut {NoStop}%
\bibitem [{\citenamefont {Knee}\ \emph {et~al.}(2016)\citenamefont {Knee},
  \citenamefont {Kakuyanagi}, \citenamefont {Yeh}, \citenamefont {Matsuzaki},
  \citenamefont {Toida}, \citenamefont {Yamaguchi}, \citenamefont {Saito},
  \citenamefont {Leggett},\ and\ \citenamefont {Munro}}]{knee2016strict}%
  \BibitemOpen
  \bibfield  {author} {\bibinfo {author} {\bibfnamefont {G.~C.}\ \bibnamefont
  {Knee}}, \bibinfo {author} {\bibfnamefont {K.}~\bibnamefont {Kakuyanagi}},
  \bibinfo {author} {\bibfnamefont {M.-C.}\ \bibnamefont {Yeh}}, \bibinfo
  {author} {\bibfnamefont {Y.}~\bibnamefont {Matsuzaki}}, \bibinfo {author}
  {\bibfnamefont {H.}~\bibnamefont {Toida}}, \bibinfo {author} {\bibfnamefont
  {H.}~\bibnamefont {Yamaguchi}}, \bibinfo {author} {\bibfnamefont
  {S.}~\bibnamefont {Saito}}, \bibinfo {author} {\bibfnamefont {A.~J.}\
  \bibnamefont {Leggett}},\ and\ \bibinfo {author} {\bibfnamefont {W.~J.}\
  \bibnamefont {Munro}},\ }\href@noop {} {\bibfield  {journal} {\bibinfo
  {journal} {Nature communications}\ }\textbf {\bibinfo {volume} {7}},\
  \bibinfo {pages} {1} (\bibinfo {year} {2016})}\BibitemShut {NoStop}%
\bibitem [{\citenamefont {Goggin}\ \emph {et~al.}(2011)\citenamefont {Goggin},
  \citenamefont {Almeida}, \citenamefont {Barbieri}, \citenamefont {Lanyon},
  \citenamefont {O’brien}, \citenamefont {White},\ and\ \citenamefont
  {Pryde}}]{goggin2011violation}%
  \BibitemOpen
  \bibfield  {author} {\bibinfo {author} {\bibfnamefont {M.}~\bibnamefont
  {Goggin}}, \bibinfo {author} {\bibfnamefont {M.}~\bibnamefont {Almeida}},
  \bibinfo {author} {\bibfnamefont {M.}~\bibnamefont {Barbieri}}, \bibinfo
  {author} {\bibfnamefont {B.}~\bibnamefont {Lanyon}}, \bibinfo {author}
  {\bibfnamefont {J.}~\bibnamefont {O’brien}}, \bibinfo {author}
  {\bibfnamefont {A.}~\bibnamefont {White}},\ and\ \bibinfo {author}
  {\bibfnamefont {G.}~\bibnamefont {Pryde}},\ }\href@noop {} {\bibfield
  {journal} {\bibinfo  {journal} {Proceedings of the National Academy of
  Sciences}\ }\textbf {\bibinfo {volume} {108}},\ \bibinfo {pages} {1256}
  (\bibinfo {year} {2011})}\BibitemShut {NoStop}%
\bibitem [{\citenamefont {Formaggio}\ \emph {et~al.}(2016)\citenamefont
  {Formaggio}, \citenamefont {Kaiser}, \citenamefont {Murskyj},\ and\
  \citenamefont {Weiss}}]{formaggio2016violation}%
  \BibitemOpen
  \bibfield  {author} {\bibinfo {author} {\bibfnamefont {J.}~\bibnamefont
  {Formaggio}}, \bibinfo {author} {\bibfnamefont {D.}~\bibnamefont {Kaiser}},
  \bibinfo {author} {\bibfnamefont {M.}~\bibnamefont {Murskyj}},\ and\ \bibinfo
  {author} {\bibfnamefont {T.}~\bibnamefont {Weiss}},\ }\href@noop {}
  {\bibfield  {journal} {\bibinfo  {journal} {Physical review letters}\
  }\textbf {\bibinfo {volume} {117}},\ \bibinfo {pages} {050402} (\bibinfo
  {year} {2016})}\BibitemShut {NoStop}%
\bibitem [{\citenamefont {Lambert}\ \emph {et~al.}(2016)\citenamefont
  {Lambert}, \citenamefont {Debnath}, \citenamefont {Kockum}, \citenamefont
  {Knee}, \citenamefont {Munro},\ and\ \citenamefont
  {Nori}}]{lambert2016leggett}%
  \BibitemOpen
  \bibfield  {author} {\bibinfo {author} {\bibfnamefont {N.}~\bibnamefont
  {Lambert}}, \bibinfo {author} {\bibfnamefont {K.}~\bibnamefont {Debnath}},
  \bibinfo {author} {\bibfnamefont {A.~F.}\ \bibnamefont {Kockum}}, \bibinfo
  {author} {\bibfnamefont {G.~C.}\ \bibnamefont {Knee}}, \bibinfo {author}
  {\bibfnamefont {W.~J.}\ \bibnamefont {Munro}},\ and\ \bibinfo {author}
  {\bibfnamefont {F.}~\bibnamefont {Nori}},\ }\href@noop {} {\bibfield
  {journal} {\bibinfo  {journal} {Physical Review A}\ }\textbf {\bibinfo
  {volume} {94}},\ \bibinfo {pages} {012105} (\bibinfo {year}
  {2016})}\BibitemShut {NoStop}%
\bibitem [{\citenamefont {Zhou}\ \emph {et~al.}(2015)\citenamefont {Zhou},
  \citenamefont {Huelga}, \citenamefont {Li},\ and\ \citenamefont
  {Guo}}]{zhou2015experimental}%
  \BibitemOpen
  \bibfield  {author} {\bibinfo {author} {\bibfnamefont {Z.-Q.}\ \bibnamefont
  {Zhou}}, \bibinfo {author} {\bibfnamefont {S.~F.}\ \bibnamefont {Huelga}},
  \bibinfo {author} {\bibfnamefont {C.-F.}\ \bibnamefont {Li}},\ and\ \bibinfo
  {author} {\bibfnamefont {G.-C.}\ \bibnamefont {Guo}},\ }\href@noop {}
  {\bibfield  {journal} {\bibinfo  {journal} {Physical review letters}\
  }\textbf {\bibinfo {volume} {115}},\ \bibinfo {pages} {113002} (\bibinfo
  {year} {2015})}\BibitemShut {NoStop}%
\bibitem [{\citenamefont {Ku}\ \emph {et~al.}(2020)\citenamefont {Ku},
  \citenamefont {Lambert}, \citenamefont {Chan}, \citenamefont {Emary},
  \citenamefont {Chen},\ and\ \citenamefont {Nori}}]{ku2020experimental}%
  \BibitemOpen
  \bibfield  {author} {\bibinfo {author} {\bibfnamefont {H.-Y.}\ \bibnamefont
  {Ku}}, \bibinfo {author} {\bibfnamefont {N.}~\bibnamefont {Lambert}},
  \bibinfo {author} {\bibfnamefont {F.-J.}\ \bibnamefont {Chan}}, \bibinfo
  {author} {\bibfnamefont {C.}~\bibnamefont {Emary}}, \bibinfo {author}
  {\bibfnamefont {Y.-N.}\ \bibnamefont {Chen}},\ and\ \bibinfo {author}
  {\bibfnamefont {F.}~\bibnamefont {Nori}},\ }\href@noop {} {\bibfield
  {journal} {\bibinfo  {journal} {npj Quantum Information}\ }\textbf {\bibinfo
  {volume} {6}},\ \bibinfo {pages} {1} (\bibinfo {year} {2020})}\BibitemShut
  {NoStop}%
\bibitem [{\citenamefont {Majidy}\ \emph {et~al.}(2019)\citenamefont {Majidy},
  \citenamefont {Katiyar}, \citenamefont {Anikeeva}, \citenamefont
  {Halliwell},\ and\ \citenamefont {Laflamme}}]{majidy2019exploration}%
  \BibitemOpen
  \bibfield  {author} {\bibinfo {author} {\bibfnamefont {S.-S.}\ \bibnamefont
  {Majidy}}, \bibinfo {author} {\bibfnamefont {H.}~\bibnamefont {Katiyar}},
  \bibinfo {author} {\bibfnamefont {G.}~\bibnamefont {Anikeeva}}, \bibinfo
  {author} {\bibfnamefont {J.~J.}\ \bibnamefont {Halliwell}},\ and\ \bibinfo
  {author} {\bibfnamefont {R.}~\bibnamefont {Laflamme}},\ }\href@noop {}
  {\bibfield  {journal} {\bibinfo  {journal} {Physical Review A}\ }\textbf
  {\bibinfo {volume} {100}},\ \bibinfo {pages} {042325} (\bibinfo {year}
  {2019})}\BibitemShut {NoStop}%
\bibitem [{\citenamefont
  {Halliwell}(2019{\natexlab{a}})}]{halliwell2019necessary}%
  \BibitemOpen
  \bibfield  {author} {\bibinfo {author} {\bibfnamefont {J.~J.}\ \bibnamefont
  {Halliwell}},\ }in\ \href@noop {} {\emph {\bibinfo {booktitle} {Journal of
  Physics: Conference Series}}},\ Vol.\ \bibinfo {volume} {1275}\ (\bibinfo
  {organization} {IOP Publishing},\ \bibinfo {year} {2019})\ p.\ \bibinfo
  {pages} {012008}\BibitemShut {NoStop}%
\bibitem [{\citenamefont {Halliwell}(2017)}]{halliwell2017comparing}%
  \BibitemOpen
  \bibfield  {author} {\bibinfo {author} {\bibfnamefont {J.~J.}\ \bibnamefont
  {Halliwell}},\ }\href@noop {} {\bibfield  {journal} {\bibinfo  {journal}
  {Physical Review A}\ }\textbf {\bibinfo {volume} {96}},\ \bibinfo {pages}
  {012121} (\bibinfo {year} {2017})}\BibitemShut {NoStop}%
\bibitem [{\citenamefont
  {Halliwell}(2016{\natexlab{a}})}]{halliwell2016leggett}%
  \BibitemOpen
  \bibfield  {author} {\bibinfo {author} {\bibfnamefont {J.~J.}\ \bibnamefont
  {Halliwell}},\ }\href@noop {} {\bibfield  {journal} {\bibinfo  {journal}
  {Physical Review A}\ }\textbf {\bibinfo {volume} {93}},\ \bibinfo {pages}
  {022123} (\bibinfo {year} {2016}{\natexlab{a}})}\BibitemShut {NoStop}%
\bibitem [{\citenamefont {Lambert}\ \emph {et~al.}(2010)\citenamefont
  {Lambert}, \citenamefont {Emary}, \citenamefont {Chen},\ and\ \citenamefont
  {Nori}}]{lambert2010distinguishing}%
  \BibitemOpen
  \bibfield  {author} {\bibinfo {author} {\bibfnamefont {N.}~\bibnamefont
  {Lambert}}, \bibinfo {author} {\bibfnamefont {C.}~\bibnamefont {Emary}},
  \bibinfo {author} {\bibfnamefont {Y.-N.}\ \bibnamefont {Chen}},\ and\
  \bibinfo {author} {\bibfnamefont {F.}~\bibnamefont {Nori}},\ }\href@noop {}
  {\bibfield  {journal} {\bibinfo  {journal} {Physical review letters}\
  }\textbf {\bibinfo {volume} {105}},\ \bibinfo {pages} {176801} (\bibinfo
  {year} {2010})}\BibitemShut {NoStop}%
\bibitem [{\citenamefont {Li}\ \emph {et~al.}(2012)\citenamefont {Li},
  \citenamefont {Lambert}, \citenamefont {Chen}, \citenamefont {Chen},\ and\
  \citenamefont {Nori}}]{li2012witnessing}%
  \BibitemOpen
  \bibfield  {author} {\bibinfo {author} {\bibfnamefont {C.-M.}\ \bibnamefont
  {Li}}, \bibinfo {author} {\bibfnamefont {N.}~\bibnamefont {Lambert}},
  \bibinfo {author} {\bibfnamefont {Y.-N.}\ \bibnamefont {Chen}}, \bibinfo
  {author} {\bibfnamefont {G.-Y.}\ \bibnamefont {Chen}},\ and\ \bibinfo
  {author} {\bibfnamefont {F.}~\bibnamefont {Nori}},\ }\href@noop {} {\bibfield
   {journal} {\bibinfo  {journal} {Scientific reports}\ }\textbf {\bibinfo
  {volume} {2}},\ \bibinfo {pages} {885} (\bibinfo {year} {2012})}\BibitemShut
  {NoStop}%
\bibitem [{\citenamefont {Lambert}\ \emph {et~al.}(2013)\citenamefont
  {Lambert}, \citenamefont {Chen}, \citenamefont {Cheng}, \citenamefont {Li},
  \citenamefont {Chen},\ and\ \citenamefont {Nori}}]{lambert2013quantum}%
  \BibitemOpen
  \bibfield  {author} {\bibinfo {author} {\bibfnamefont {N.}~\bibnamefont
  {Lambert}}, \bibinfo {author} {\bibfnamefont {Y.-N.}\ \bibnamefont {Chen}},
  \bibinfo {author} {\bibfnamefont {Y.-C.}\ \bibnamefont {Cheng}}, \bibinfo
  {author} {\bibfnamefont {C.-M.}\ \bibnamefont {Li}}, \bibinfo {author}
  {\bibfnamefont {G.-Y.}\ \bibnamefont {Chen}},\ and\ \bibinfo {author}
  {\bibfnamefont {F.}~\bibnamefont {Nori}},\ }\href@noop {} {\bibfield
  {journal} {\bibinfo  {journal} {Nature Physics}\ }\textbf {\bibinfo {volume}
  {9}},\ \bibinfo {pages} {10} (\bibinfo {year} {2013})}\BibitemShut {NoStop}%
\bibitem [{\citenamefont {Friedenberger}\ and\ \citenamefont
  {Lutz}(2017)}]{friedenberger2017quantum}%
  \BibitemOpen
  \bibfield  {author} {\bibinfo {author} {\bibfnamefont {A.}~\bibnamefont
  {Friedenberger}}\ and\ \bibinfo {author} {\bibfnamefont {E.}~\bibnamefont
  {Lutz}},\ }\href@noop {} {\bibfield  {journal} {\bibinfo  {journal} {EPL
  (Europhysics Letters)}\ }\textbf {\bibinfo {volume} {120}},\ \bibinfo {pages}
  {10002} (\bibinfo {year} {2017})}\BibitemShut {NoStop}%
\bibitem [{\citenamefont
  {Halliwell}(2019{\natexlab{b}})}]{halliwell2019leggett}%
  \BibitemOpen
  \bibfield  {author} {\bibinfo {author} {\bibfnamefont {J.~J.}\ \bibnamefont
  {Halliwell}},\ }\href@noop {} {\bibfield  {journal} {\bibinfo  {journal}
  {Physical Review A}\ }\textbf {\bibinfo {volume} {99}},\ \bibinfo {pages}
  {022119} (\bibinfo {year} {2019}{\natexlab{b}})}\BibitemShut {NoStop}%
\bibitem [{\citenamefont {Avis}\ \emph {et~al.}(2010)\citenamefont {Avis},
  \citenamefont {Hayden},\ and\ \citenamefont {Wilde}}]{avis2010leggett}%
  \BibitemOpen
  \bibfield  {author} {\bibinfo {author} {\bibfnamefont {D.}~\bibnamefont
  {Avis}}, \bibinfo {author} {\bibfnamefont {P.}~\bibnamefont {Hayden}},\ and\
  \bibinfo {author} {\bibfnamefont {M.~M.}\ \bibnamefont {Wilde}},\ }\href@noop
  {} {\bibfield  {journal} {\bibinfo  {journal} {Physical Review A}\ }\textbf
  {\bibinfo {volume} {82}},\ \bibinfo {pages} {030102(R)} (\bibinfo {year}
  {2010})}\BibitemShut {NoStop}%
\bibitem [{\citenamefont {Halliwell}\ and\ \citenamefont
  {Mawby}(2019)}]{halliwell2019fine}%
  \BibitemOpen
  \bibfield  {author} {\bibinfo {author} {\bibfnamefont {J.~J.}\ \bibnamefont
  {Halliwell}}\ and\ \bibinfo {author} {\bibfnamefont {C.}~\bibnamefont
  {Mawby}},\ }\href@noop {} {\bibfield  {journal} {\bibinfo  {journal}
  {Physical Review A}\ }\textbf {\bibinfo {volume} {100}},\ \bibinfo {pages}
  {042103} (\bibinfo {year} {2019})}\BibitemShut {NoStop}%
\bibitem [{\citenamefont {Halliwell}\ and\ \citenamefont
  {Mawby}(2020)}]{halliwell2020conditions}%
  \BibitemOpen
  \bibfield  {author} {\bibinfo {author} {\bibfnamefont {J.~J.}\ \bibnamefont
  {Halliwell}}\ and\ \bibinfo {author} {\bibfnamefont {C.}~\bibnamefont
  {Mawby}},\ }\href {https://doi.org/10.1103/PhysRevA.102.012209} {\bibfield
  {journal} {\bibinfo  {journal} {Physical Review A}\ }\textbf {\bibinfo
  {volume} {102}},\ \bibinfo {pages} {012209} (\bibinfo {year}
  {2020})}\BibitemShut {NoStop}%
\bibitem [{\citenamefont {Kauten}\ \emph {et~al.}(2017)\citenamefont {Kauten},
  \citenamefont {Keil}, \citenamefont {Kaufmann}, \citenamefont {Pressl},
  \citenamefont {Brukner},\ and\ \citenamefont {Weihs}}]{kauten2017obtaining}%
  \BibitemOpen
  \bibfield  {author} {\bibinfo {author} {\bibfnamefont {T.}~\bibnamefont
  {Kauten}}, \bibinfo {author} {\bibfnamefont {R.}~\bibnamefont {Keil}},
  \bibinfo {author} {\bibfnamefont {T.}~\bibnamefont {Kaufmann}}, \bibinfo
  {author} {\bibfnamefont {B.}~\bibnamefont {Pressl}}, \bibinfo {author}
  {\bibfnamefont {{\v{C}}.}~\bibnamefont {Brukner}},\ and\ \bibinfo {author}
  {\bibfnamefont {G.}~\bibnamefont {Weihs}},\ }\href@noop {} {\bibfield
  {journal} {\bibinfo  {journal} {New Journal of Physics}\ }\textbf {\bibinfo
  {volume} {19}},\ \bibinfo {pages} {033017} (\bibinfo {year}
  {2017})}\BibitemShut {NoStop}%
\bibitem [{\citenamefont {Hofer}(2017)}]{hofer2017quasi}%
  \BibitemOpen
  \bibfield  {author} {\bibinfo {author} {\bibfnamefont {P.~P.}\ \bibnamefont
  {Hofer}},\ }\href@noop {} {\bibfield  {journal} {\bibinfo  {journal}
  {Quantum}\ }\textbf {\bibinfo {volume} {1}},\ \bibinfo {pages} {32} (\bibinfo
  {year} {2017})}\BibitemShut {NoStop}%
\bibitem [{\citenamefont {Dressel}(2015)}]{dressel2015weak}%
  \BibitemOpen
  \bibfield  {author} {\bibinfo {author} {\bibfnamefont {J.}~\bibnamefont
  {Dressel}},\ }\href@noop {} {\bibfield  {journal} {\bibinfo  {journal}
  {Physical Review A}\ }\textbf {\bibinfo {volume} {91}},\ \bibinfo {pages}
  {032116} (\bibinfo {year} {2015})}\BibitemShut {NoStop}%
\bibitem [{\citenamefont {Pan}(2020)}]{pan2020interference}%
  \BibitemOpen
  \bibfield  {author} {\bibinfo {author} {\bibfnamefont {A.}~\bibnamefont
  {Pan}},\ }\href@noop {} {\bibfield  {journal} {\bibinfo  {journal} {Physical
  Review A}\ }\textbf {\bibinfo {volume} {102}},\ \bibinfo {pages} {032206}
  (\bibinfo {year} {2020})}\BibitemShut {NoStop}%
\bibitem [{\citenamefont {Kofler}\ and\ \citenamefont
  {Brukner}(2013)}]{kofler2013condition}%
  \BibitemOpen
  \bibfield  {author} {\bibinfo {author} {\bibfnamefont {J.}~\bibnamefont
  {Kofler}}\ and\ \bibinfo {author} {\bibfnamefont {{\v{C}}.}~\bibnamefont
  {Brukner}},\ }\href@noop {} {\bibfield  {journal} {\bibinfo  {journal}
  {Physical Review A}\ }\textbf {\bibinfo {volume} {87}},\ \bibinfo {pages}
  {052115} (\bibinfo {year} {2013})}\BibitemShut {NoStop}%
\bibitem [{\citenamefont {Clemente}\ and\ \citenamefont
  {Kofler}(2015)}]{clemente2015necessary}%
  \BibitemOpen
  \bibfield  {author} {\bibinfo {author} {\bibfnamefont {L.}~\bibnamefont
  {Clemente}}\ and\ \bibinfo {author} {\bibfnamefont {J.}~\bibnamefont
  {Kofler}},\ }\href@noop {} {\bibfield  {journal} {\bibinfo  {journal}
  {Physical Review A}\ }\textbf {\bibinfo {volume} {91}},\ \bibinfo {pages}
  {062103} (\bibinfo {year} {2015})}\BibitemShut {NoStop}%
\bibitem [{\citenamefont {Deza}\ and\ \citenamefont
  {Laurent}(2009)}]{deza2009geometry}%
  \BibitemOpen
  \bibfield  {author} {\bibinfo {author} {\bibfnamefont {M.~M.}\ \bibnamefont
  {Deza}}\ and\ \bibinfo {author} {\bibfnamefont {M.}~\bibnamefont {Laurent}},\
  }\href@noop {} {\emph {\bibinfo {title} {Geometry of cuts and metrics}}},\
  Vol.~\bibinfo {volume} {15}\ (\bibinfo  {publisher} {Springer},\ \bibinfo
  {year} {2009})\BibitemShut {NoStop}%
\bibitem [{\citenamefont {Avis}(1978)}]{avis1978some}%
  \BibitemOpen
  \bibfield  {author} {\bibinfo {author} {\bibfnamefont {D.~M.}\ \bibnamefont
  {Avis}},\ }\emph {\bibinfo {title} {Some polyhedral cones related to metric
  spaces}},\ \href@noop {} {Ph.D. thesis},\ \bibinfo  {school} {Stanford
  University} (\bibinfo {year} {1978})\BibitemShut {NoStop}%
\bibitem [{sup(2021)}]{supplementary}%
  \BibitemOpen
  \href@noop {} {\bibinfo {title} {See {S}upplemental {M}aterial at [url will
  be inserted by publisher] for the full details of the numerical searches done
  in this work.}} (\bibinfo {year} {2021})\BibitemShut {NoStop}%
\bibitem [{\citenamefont {Bose}\ \emph {et~al.}(2018)\citenamefont {Bose},
  \citenamefont {Home},\ and\ \citenamefont {Mal}}]{bose2018nonclassicality}%
  \BibitemOpen
  \bibfield  {author} {\bibinfo {author} {\bibfnamefont {S.}~\bibnamefont
  {Bose}}, \bibinfo {author} {\bibfnamefont {D.}~\bibnamefont {Home}},\ and\
  \bibinfo {author} {\bibfnamefont {S.}~\bibnamefont {Mal}},\ }\href@noop {}
  {\bibfield  {journal} {\bibinfo  {journal} {Physical review letters}\
  }\textbf {\bibinfo {volume} {120}},\ \bibinfo {pages} {210402} (\bibinfo
  {year} {2018})}\BibitemShut {NoStop}%
\bibitem [{\citenamefont {Halliwell}\ \emph {et~al.}(2021)\citenamefont
  {Halliwell}, \citenamefont {Bhatnagar}, \citenamefont {Ireland},
  \citenamefont {Nadeem},\ and\ \citenamefont
  {Wimalaweera}}]{halliwell2021leggett}%
  \BibitemOpen
  \bibfield  {author} {\bibinfo {author} {\bibfnamefont {J.}~\bibnamefont
  {Halliwell}}, \bibinfo {author} {\bibfnamefont {A.}~\bibnamefont
  {Bhatnagar}}, \bibinfo {author} {\bibfnamefont {E.}~\bibnamefont {Ireland}},
  \bibinfo {author} {\bibfnamefont {H.}~\bibnamefont {Nadeem}},\ and\ \bibinfo
  {author} {\bibfnamefont {V.}~\bibnamefont {Wimalaweera}},\ }\href@noop {}
  {\bibfield  {journal} {\bibinfo  {journal} {Physical Review A}\ }\textbf
  {\bibinfo {volume} {103}},\ \bibinfo {pages} {032218} (\bibinfo {year}
  {2021})}\BibitemShut {NoStop}%
\bibitem [{\citenamefont {Montina}(2012)}]{montina2012dynamics}%
  \BibitemOpen
  \bibfield  {author} {\bibinfo {author} {\bibfnamefont {A.}~\bibnamefont
  {Montina}},\ }\href@noop {} {\bibfield  {journal} {\bibinfo  {journal}
  {Physical review letters}\ }\textbf {\bibinfo {volume} {108}},\ \bibinfo
  {pages} {160501} (\bibinfo {year} {2012})}\BibitemShut {NoStop}%
\bibitem [{\citenamefont {Yearsley}(2013)}]{yearsley2013leggett}%
  \BibitemOpen
  \bibfield  {author} {\bibinfo {author} {\bibfnamefont {J.}~\bibnamefont
  {Yearsley}},\ }\href@noop {} {\bibfield  {journal} {\bibinfo  {journal}
  {arXiv preprint arXiv:1310.2149}\ } (\bibinfo {year} {2013})}\BibitemShut
  {NoStop}%
\bibitem [{\citenamefont
  {Halliwell}(2016{\natexlab{b}})}]{halliwell2016leggett_ctvm}%
  \BibitemOpen
  \bibfield  {author} {\bibinfo {author} {\bibfnamefont {J.}~\bibnamefont
  {Halliwell}},\ }\href@noop {} {\bibfield  {journal} {\bibinfo  {journal}
  {Physical Review A}\ }\textbf {\bibinfo {volume} {94}},\ \bibinfo {pages}
  {052114} (\bibinfo {year} {2016}{\natexlab{b}})}\BibitemShut {NoStop}%
\bibitem [{\citenamefont {Pan}\ \emph {et~al.}(2018)\citenamefont {Pan},
  \citenamefont {Qutubuddin},\ and\ \citenamefont {Kumari}}]{pan2018quantum}%
  \BibitemOpen
  \bibfield  {author} {\bibinfo {author} {\bibfnamefont {A.~K.}\ \bibnamefont
  {Pan}}, \bibinfo {author} {\bibfnamefont {M.}~\bibnamefont {Qutubuddin}},\
  and\ \bibinfo {author} {\bibfnamefont {S.}~\bibnamefont {Kumari}},\
  }\href@noop {} {\bibfield  {journal} {\bibinfo  {journal} {Physical Review
  A}\ }\textbf {\bibinfo {volume} {98}},\ \bibinfo {pages} {062115} (\bibinfo
  {year} {2018})}\BibitemShut {NoStop}%
\bibitem [{\citenamefont {Bechtold}\ \emph {et~al.}(2016)\citenamefont
  {Bechtold}, \citenamefont {Li}, \citenamefont {M{\"u}ller}, \citenamefont
  {Simmet}, \citenamefont {Ardelt}, \citenamefont {Finley},\ and\ \citenamefont
  {Sinitsyn}}]{bechtold2016quantum}%
  \BibitemOpen
  \bibfield  {author} {\bibinfo {author} {\bibfnamefont {A.}~\bibnamefont
  {Bechtold}}, \bibinfo {author} {\bibfnamefont {F.}~\bibnamefont {Li}},
  \bibinfo {author} {\bibfnamefont {K.}~\bibnamefont {M{\"u}ller}}, \bibinfo
  {author} {\bibfnamefont {T.}~\bibnamefont {Simmet}}, \bibinfo {author}
  {\bibfnamefont {P.-L.}\ \bibnamefont {Ardelt}}, \bibinfo {author}
  {\bibfnamefont {J.~J.}\ \bibnamefont {Finley}},\ and\ \bibinfo {author}
  {\bibfnamefont {N.~A.}\ \bibnamefont {Sinitsyn}},\ }\href@noop {} {\bibfield
  {journal} {\bibinfo  {journal} {Physical Review Letters}\ }\textbf {\bibinfo
  {volume} {117}},\ \bibinfo {pages} {027402} (\bibinfo {year}
  {2016})}\BibitemShut {NoStop}%
\bibitem [{\citenamefont {Budroni}\ and\ \citenamefont
  {Emary}(2014)}]{budroni2014temporal}%
  \BibitemOpen
  \bibfield  {author} {\bibinfo {author} {\bibfnamefont {C.}~\bibnamefont
  {Budroni}}\ and\ \bibinfo {author} {\bibfnamefont {C.}~\bibnamefont
  {Emary}},\ }\href@noop {} {\bibfield  {journal} {\bibinfo  {journal}
  {Physical review letters}\ }\textbf {\bibinfo {volume} {113}},\ \bibinfo
  {pages} {050401} (\bibinfo {year} {2014})}\BibitemShut {NoStop}%
\bibitem [{\citenamefont {Daki{\'c}}\ \emph {et~al.}(2014)\citenamefont
  {Daki{\'c}}, \citenamefont {Paterek},\ and\ \citenamefont
  {Brukner}}]{dakic2014density}%
  \BibitemOpen
  \bibfield  {author} {\bibinfo {author} {\bibfnamefont {B.}~\bibnamefont
  {Daki{\'c}}}, \bibinfo {author} {\bibfnamefont {T.}~\bibnamefont {Paterek}},\
  and\ \bibinfo {author} {\bibfnamefont {{\v{C}}.}~\bibnamefont {Brukner}},\
  }\href@noop {} {\bibfield  {journal} {\bibinfo  {journal} {New Journal of
  Physics}\ }\textbf {\bibinfo {volume} {16}},\ \bibinfo {pages} {023028}
  (\bibinfo {year} {2014})}\BibitemShut {NoStop}%
\bibitem [{\citenamefont {Wang}\ \emph {et~al.}(2017)\citenamefont {Wang},
  \citenamefont {Emary}, \citenamefont {Zhan}, \citenamefont {Bian},
  \citenamefont {Li},\ and\ \citenamefont {Xue}}]{wang2017enhanced}%
  \BibitemOpen
  \bibfield  {author} {\bibinfo {author} {\bibfnamefont {K.}~\bibnamefont
  {Wang}}, \bibinfo {author} {\bibfnamefont {C.}~\bibnamefont {Emary}},
  \bibinfo {author} {\bibfnamefont {X.}~\bibnamefont {Zhan}}, \bibinfo {author}
  {\bibfnamefont {Z.}~\bibnamefont {Bian}}, \bibinfo {author} {\bibfnamefont
  {J.}~\bibnamefont {Li}},\ and\ \bibinfo {author} {\bibfnamefont
  {P.}~\bibnamefont {Xue}},\ }\href@noop {} {\bibfield  {journal} {\bibinfo
  {journal} {Optics express}\ }\textbf {\bibinfo {volume} {25}},\ \bibinfo
  {pages} {31462} (\bibinfo {year} {2017})}\BibitemShut {NoStop}%
\bibitem [{\citenamefont {Kumari}\ \emph {et~al.}(2018)\citenamefont {Kumari},
  \citenamefont {Qutubuddin},\ and\ \citenamefont {Pan}}]{kumari2018violation}%
  \BibitemOpen
  \bibfield  {author} {\bibinfo {author} {\bibfnamefont {A.}~\bibnamefont
  {Kumari}}, \bibinfo {author} {\bibfnamefont {M.}~\bibnamefont {Qutubuddin}},\
  and\ \bibinfo {author} {\bibfnamefont {A.~K.}\ \bibnamefont {Pan}},\
  }\href@noop {} {\bibfield  {journal} {\bibinfo  {journal} {Physical Review
  A}\ }\textbf {\bibinfo {volume} {98}},\ \bibinfo {pages} {042135} (\bibinfo
  {year} {2018})}\BibitemShut {NoStop}%
\bibitem [{\citenamefont {Fritz}(2010)}]{fritz2010quantum}%
  \BibitemOpen
  \bibfield  {author} {\bibinfo {author} {\bibfnamefont {T.}~\bibnamefont
  {Fritz}},\ }\href@noop {} {\bibfield  {journal} {\bibinfo  {journal} {New
  Journal of Physics}\ }\textbf {\bibinfo {volume} {12}},\ \bibinfo {pages}
  {083055} (\bibinfo {year} {2010})}\BibitemShut {NoStop}%
\bibitem [{\citenamefont {Athalye}\ \emph {et~al.}(2011)\citenamefont
  {Athalye}, \citenamefont {Roy},\ and\ \citenamefont
  {Mahesh}}]{athalye2011investigation}%
  \BibitemOpen
  \bibfield  {author} {\bibinfo {author} {\bibfnamefont {V.}~\bibnamefont
  {Athalye}}, \bibinfo {author} {\bibfnamefont {S.~S.}\ \bibnamefont {Roy}},\
  and\ \bibinfo {author} {\bibfnamefont {T.}~\bibnamefont {Mahesh}},\
  }\href@noop {} {\bibfield  {journal} {\bibinfo  {journal} {Physical review
  letters}\ }\textbf {\bibinfo {volume} {107}},\ \bibinfo {pages} {130402}
  (\bibinfo {year} {2011})}\BibitemShut {NoStop}%
\bibitem [{\citenamefont {Majidy}(2019)}]{majidy2019violation}%
  \BibitemOpen
  \bibfield  {author} {\bibinfo {author} {\bibfnamefont {S.-S.}\ \bibnamefont
  {Majidy}},\ }\emph {\bibinfo {title} {Violation of an augmented set of
  Leggett-Garg inequalities and the implementation of a continuous in time
  velocity measurement}},\ \href@noop {} {Master's thesis},\ \bibinfo  {school}
  {University of Waterloo} (\bibinfo {year} {2019})\BibitemShut {NoStop}%
\bibitem [{\citenamefont {Katiyar}\ \emph {et~al.}(2017)\citenamefont
  {Katiyar}, \citenamefont {Brodutch}, \citenamefont {Lu},\ and\ \citenamefont
  {Laflamme}}]{katiyar2017experimental}%
  \BibitemOpen
  \bibfield  {author} {\bibinfo {author} {\bibfnamefont {H.}~\bibnamefont
  {Katiyar}}, \bibinfo {author} {\bibfnamefont {A.}~\bibnamefont {Brodutch}},
  \bibinfo {author} {\bibfnamefont {D.}~\bibnamefont {Lu}},\ and\ \bibinfo
  {author} {\bibfnamefont {R.}~\bibnamefont {Laflamme}},\ }\href@noop {}
  {\bibfield  {journal} {\bibinfo  {journal} {New Journal of Physics}\ }\textbf
  {\bibinfo {volume} {19}},\ \bibinfo {pages} {023033} (\bibinfo {year}
  {2017})}\BibitemShut {NoStop}%
\bibitem [{\citenamefont {Levitt}(2013)}]{levitt2013spin}%
  \BibitemOpen
  \bibfield  {author} {\bibinfo {author} {\bibfnamefont {M.~H.}\ \bibnamefont
  {Levitt}},\ }\href@noop {} {\emph {\bibinfo {title} {Spin dynamics: basics of
  nuclear magnetic resonance}}}\ (\bibinfo  {publisher} {John Wiley \& Sons},\
  \bibinfo {year} {2013})\BibitemShut {NoStop}%
\bibitem [{\citenamefont {Bruzewicz}\ \emph {et~al.}(2019)\citenamefont
  {Bruzewicz}, \citenamefont {Chiaverini}, \citenamefont {McConnell},\ and\
  \citenamefont {Sage}}]{bruzewicz2019trapped}%
  \BibitemOpen
  \bibfield  {author} {\bibinfo {author} {\bibfnamefont {C.~D.}\ \bibnamefont
  {Bruzewicz}}, \bibinfo {author} {\bibfnamefont {J.}~\bibnamefont
  {Chiaverini}}, \bibinfo {author} {\bibfnamefont {R.}~\bibnamefont
  {McConnell}},\ and\ \bibinfo {author} {\bibfnamefont {J.~M.}\ \bibnamefont
  {Sage}},\ }\href@noop {} {\bibfield  {journal} {\bibinfo  {journal} {Applied
  Physics Reviews}\ }\textbf {\bibinfo {volume} {6}},\ \bibinfo {pages}
  {021314} (\bibinfo {year} {2019})}\BibitemShut {NoStop}%
\bibitem [{\citenamefont {Blais}\ \emph {et~al.}(2020)\citenamefont {Blais},
  \citenamefont {Girvin},\ and\ \citenamefont {Oliver}}]{blais2020quantum}%
  \BibitemOpen
  \bibfield  {author} {\bibinfo {author} {\bibfnamefont {A.}~\bibnamefont
  {Blais}}, \bibinfo {author} {\bibfnamefont {S.~M.}\ \bibnamefont {Girvin}},\
  and\ \bibinfo {author} {\bibfnamefont {W.~D.}\ \bibnamefont {Oliver}},\
  }\href@noop {} {\bibfield  {journal} {\bibinfo  {journal} {Nature Physics}\
  }\textbf {\bibinfo {volume} {16}},\ \bibinfo {pages} {247} (\bibinfo {year}
  {2020})}\BibitemShut {NoStop}%
\bibitem [{\citenamefont {Lunghi}\ \emph {et~al.}(2015)\citenamefont {Lunghi},
  \citenamefont {Brask}, \citenamefont {Lim}, \citenamefont {Lavigne},
  \citenamefont {Bowles}, \citenamefont {Martin}, \citenamefont {Zbinden},\
  and\ \citenamefont {Brunner}}]{lunghi2015self}%
  \BibitemOpen
  \bibfield  {author} {\bibinfo {author} {\bibfnamefont {T.}~\bibnamefont
  {Lunghi}}, \bibinfo {author} {\bibfnamefont {J.~B.}\ \bibnamefont {Brask}},
  \bibinfo {author} {\bibfnamefont {C.~C.~W.}\ \bibnamefont {Lim}}, \bibinfo
  {author} {\bibfnamefont {Q.}~\bibnamefont {Lavigne}}, \bibinfo {author}
  {\bibfnamefont {J.}~\bibnamefont {Bowles}}, \bibinfo {author} {\bibfnamefont
  {A.}~\bibnamefont {Martin}}, \bibinfo {author} {\bibfnamefont
  {H.}~\bibnamefont {Zbinden}},\ and\ \bibinfo {author} {\bibfnamefont
  {N.}~\bibnamefont {Brunner}},\ }\href@noop {} {\bibfield  {journal} {\bibinfo
   {journal} {Physical review letters}\ }\textbf {\bibinfo {volume} {114}},\
  \bibinfo {pages} {150501} (\bibinfo {year} {2015})}\BibitemShut {NoStop}%
\bibitem [{\citenamefont {Li}\ \emph {et~al.}(2011)\citenamefont {Li},
  \citenamefont {Yin}, \citenamefont {Wu}, \citenamefont {Zou}, \citenamefont
  {Wang}, \citenamefont {Chen}, \citenamefont {Guo},\ and\ \citenamefont
  {Han}}]{li2011semi}%
  \BibitemOpen
  \bibfield  {author} {\bibinfo {author} {\bibfnamefont {H.-W.}\ \bibnamefont
  {Li}}, \bibinfo {author} {\bibfnamefont {Z.-Q.}\ \bibnamefont {Yin}},
  \bibinfo {author} {\bibfnamefont {Y.-C.}\ \bibnamefont {Wu}}, \bibinfo
  {author} {\bibfnamefont {X.-B.}\ \bibnamefont {Zou}}, \bibinfo {author}
  {\bibfnamefont {S.}~\bibnamefont {Wang}}, \bibinfo {author} {\bibfnamefont
  {W.}~\bibnamefont {Chen}}, \bibinfo {author} {\bibfnamefont {G.-C.}\
  \bibnamefont {Guo}},\ and\ \bibinfo {author} {\bibfnamefont {Z.-F.}\
  \bibnamefont {Han}},\ }\href@noop {} {\bibfield  {journal} {\bibinfo
  {journal} {Physical Review A}\ }\textbf {\bibinfo {volume} {84}},\ \bibinfo
  {pages} {034301} (\bibinfo {year} {2011})}\BibitemShut {NoStop}%
\bibitem [{\citenamefont {Maity}\ \emph {et~al.}(2021)\citenamefont {Maity},
  \citenamefont {Mal}, \citenamefont {Jebarathinam},\ and\ \citenamefont
  {Majumdar}}]{maity2020self}%
  \BibitemOpen
  \bibfield  {author} {\bibinfo {author} {\bibfnamefont {A.~G.}\ \bibnamefont
  {Maity}}, \bibinfo {author} {\bibfnamefont {S.}~\bibnamefont {Mal}}, \bibinfo
  {author} {\bibfnamefont {C.}~\bibnamefont {Jebarathinam}},\ and\ \bibinfo
  {author} {\bibfnamefont {A.}~\bibnamefont {Majumdar}},\ }\href@noop {}
  {\bibfield  {journal} {\bibinfo  {journal} {Physical Review A}\ }\textbf
  {\bibinfo {volume} {103}},\ \bibinfo {pages} {062604} (\bibinfo {year}
  {2021})}\BibitemShut {NoStop}%
\bibitem [{\citenamefont {Berg}\ \emph {et~al.}(2009)\citenamefont {Berg},
  \citenamefont {Plimak}, \citenamefont {Polkovnikov}, \citenamefont {Olsen},
  \citenamefont {Fleischhauer},\ and\ \citenamefont
  {Schleich}}]{berg2009commuting}%
  \BibitemOpen
  \bibfield  {author} {\bibinfo {author} {\bibfnamefont {B.}~\bibnamefont
  {Berg}}, \bibinfo {author} {\bibfnamefont {L.~I.}\ \bibnamefont {Plimak}},
  \bibinfo {author} {\bibfnamefont {A.}~\bibnamefont {Polkovnikov}}, \bibinfo
  {author} {\bibfnamefont {M.}~\bibnamefont {Olsen}}, \bibinfo {author}
  {\bibfnamefont {M.}~\bibnamefont {Fleischhauer}},\ and\ \bibinfo {author}
  {\bibfnamefont {W.~P.}\ \bibnamefont {Schleich}},\ }\href@noop {} {\bibfield
  {journal} {\bibinfo  {journal} {Physical Review A}\ }\textbf {\bibinfo
  {volume} {80}},\ \bibinfo {pages} {033624} (\bibinfo {year}
  {2009})}\BibitemShut {NoStop}%
\bibitem [{\citenamefont {Menda{\v{s}}}(2006)}]{mendavs2006classification}%
  \BibitemOpen
  \bibfield  {author} {\bibinfo {author} {\bibfnamefont {I.~P.}\ \bibnamefont
  {Menda{\v{s}}}},\ }\href@noop {} {\bibfield  {journal} {\bibinfo  {journal}
  {Journal of Physics A: Mathematical and General}\ }\textbf {\bibinfo {volume}
  {39}},\ \bibinfo {pages} {11313} (\bibinfo {year} {2006})}\BibitemShut
  {NoStop}%
\end{thebibliography}%
\end{document}